\documentclass[twocolumn,numberedappendix,breaklinks=true]{aastex63}

\usepackage{natbib}
\usepackage{multirow}
\usepackage{gensymb} 
\usepackage[flushleft]{threeparttable} 
\usepackage{amsmath}
\usepackage{url}
\usepackage{multirow}
\usepackage{wasysym}

\shorttitle{GREATS Data Release}
\shortauthors{Stefanon et al.}

\begin{document}

\title{The Spitzer/IRAC Legacy over the GOODS Fields:  Full-Depth  $3.6$, $4.5$, $5.8$ and $8.0$\,$\mu$m Mosaics and Photometry for $>9000$ Galaxies at $z\sim3.5-10$ from the GOODS Re-ionization Era wide-Area Treasury from Spitzer (GREATS)}

\author{Mauro Stefanon}
\affiliation{Leiden Observatory, Leiden University, NL-2300 RA Leiden, Netherlands}

\author{Ivo Labb\'e}
\affiliation{Centre for Astrophysics and SuperComputing, Swinburne, University of Technology, Hawthorn, Victoria, 3122, Australia}

\author{Pascal A. Oesch}
\affiliation{Departement d'Astronomie, Universit\'e de Gen\'eve, 51 Ch. des Maillettes, CH-1290 Versoix, Switzerland}
\affiliation{Cosmic Dawn Center (DAWN), Niels Bohr Institute, University of Copenhagen, Jagtvej 128, K\o benhavn N, DK-2200, Denmark}

\author{Stephane de Barros}
\affiliation{Departement d'Astronomie, Universit\'e de Gen\'eve, 51 Ch. des Maillettes, CH-1290 Versoix, Switzerland}

\author{Valentino Gonzalez}
\affiliation{Departamento de Astronom\'ia, Universidad de Chile, Casilla 36-D, Santiago 7591245, Chile}
\affiliation{Centro de Astrof\'isica y Tecnologias Afines (CATA), Camino del Observatorio 1515, Las Condes, Santiago 7591245, Chile}

\author{Rychard J. Bouwens}
\affiliation{Leiden Observatory, Leiden University, NL-2300 RA Leiden, Netherlands}

\author{Marijn Franx}
\affiliation{Leiden Observatory, Leiden University, NL-2300 RA Leiden, Netherlands}

\author{Garth D. Illingworth}
\affiliation{UCO/Lick Observatory, University of California, Santa Cruz, 1156 High St, Santa Cruz, CA 95064, USA}

\author{Brad Holden}
\affiliation{UCO/Lick Observatory, University of California, Santa Cruz, 1156 High St, Santa Cruz, CA 95064, USA}

\author{Dan Magee}
\affiliation{UCO/Lick Observatory, University of California, Santa Cruz, 1156 High St, Santa Cruz, CA 95064, USA}

\author{Renske Smit}
\affiliation{Liverpool John Moores University, Liverpool, UK}

\author{Pieter van Dokkum}
\affiliation{Astronomy Department, Yale University, 52 Hillhouse Ave, New Haven, CT 06511, USA}

\email{Email: stefanon@strw.leidenuniv.nl}

\begin{abstract}
We present the deepest \textit{Spitzer}/IRAC $3.6$, $4.5$, $5.8$ and $8.0$\,$\mu$m wide-area mosaics yet over the GOODS-N and GOODS-S fields as part of the \textit{GOODS Re-ionization Era wide-Area Treasury from Spitzer} (GREATS) project. We reduced and mosaicked in a self-consistent way observations taken by the 11 different \textit{Spitzer}/IRAC programs over the two GOODS fields from 12 years of \textit{Spitzer} cryogenic and warm mission data. The cumulative depth in the $3.6$\,$\mu$m and $4.5$\,$\mu$m bands amounts to $\sim 4260$\,hr, $\sim 1220$\,hr of which are new very deep observations from the GREATS program itself. In the deepest area, the full-depth mosaics reach $\gtrsim200$\,hr over an area of $\sim100$\,arcmin$^2$, corresponding to a sensitivity of $\sim29$\,AB magnitude at $3.6$\,$\mu$m ($1\sigma$ for point sources). Archival cryogenic $5.8$\,$\mu$m and $8.0$\,$\mu$m band data (a cumulative 976 hr) are also included in the release. The mosaics are projected onto the tangential plane of CANDELS/GOODS at a $0\farcs3$\,pixel$^{-1}$ scale. This paper describes the methodology enabling, and the characteristics of, the public release of the mosaic science images, the corresponding coverage maps in the four IRAC bands and the empirical Point-Spread Functions (PSFs). These PSFs enable mitigation of the source blending effects by taking into account the complex position-dependent variation in the IRAC images. The GREATS data products are in the Infrared Science Archive (IRSA). We also release the deblended $3.6$-to-$8.0$\,$\mu$m photometry $9192$ Lyman-Break galaxies at $z\sim3.5-10$. GREATS will be the deepest mid-infrared imaging until \textit{JWST} and, as such, constitutes a major resource for characterising early galaxy assembly.
\end{abstract}


\section{Introduction}

\begin{deluxetable*}{lrlccccrrc}
\tabcolsep=0.08cm
\tablecaption{Summary of \textit{Spitzer/IRAC} datasets included in the GREATS mosaics \label{tab:programs}}
\tablehead{\colhead{Program name} & \colhead{PID\tablenotemark{a}} & \colhead{PI\tablenotemark{b}} & \colhead{Year\tablenotemark{c}} & \colhead{Max cov.\tablenotemark{d}}  & \colhead{\#} & \colhead{Total cov.\tablenotemark{f}} & \colhead{\# frames\tablenotemark{g}} & \colhead{SSC Pipeline} & \colhead{Ref.\tablenotemark{h}} \\
 & & & &  & \colhead{point.\tablenotemark{e}} & & & \colhead{version} & \\
 & & & & \colhead{[hr]} &  & \colhead{[hr]} & & & 
}
\startdata
\multicolumn{10}{c}{GOODS-N}  \\
 GOODS & $      169 $ & Dickinson\tablenotemark{$\dagger$} &$2004-2005$ & $   92.8 $ & $ 8 $ & $  278.3$ & $   5364 $ &                       S18.25.0   & [1] \\
          SEDS & $    61040 $ &     Fazio & $2010-2011$ & $   12.6 $ & $ 33 $ & $  179.1 $ & $   6947 $ &              S18.18.0/S19.1.0    & [2]\\
     S-CANDELS & $    80215 $ &     Fazio & $2012$ & $   25.3 $ & $  4 $ & $  101.2 $ & $   3944 $ &                       S19.1.0    & [5]\\
        GREATS & $    11134 $ &   Labb\'e & $2015-2016$ & $  122.6 $ & $  4 $ & $  372.6 $ & $  14628 $ &              S19.1.0 /S19.2.0    & [6]\\
       \hline
        & \multicolumn{2}{r}{Totals\tablenotemark{i}:} & & $288.9$\tablenotemark{$\ddagger$} & $49$ & $931.2$ & $30833$ & & \\
       \hline
& & & & & & & & \\ 
\multicolumn{10}{c}{GOODS-S}  \\
 GOODS & $      194 $ & Dickinson\tablenotemark{$\dagger$} & $2004$ & $   46.9 $ & $  8 $ & $  180.7$ & $   3494 $ &                       S18.25.0   & [1] \\
          UDF2 & $    30866 $ &   Bouwens\tablenotemark{$\dagger$} &$2006-2007$ &$   26.9 $ & $  1 $ & $   29.1$ & $   1098 $ &                       S18.25.0   & [3]\\
          SEDS & $    60022 $ &     Fazio & $2010-2011$ & $    8.8 $ & $ 54 $ & $  211.8 $ & $   8146 $ &              S18.18.0/S19.0.0    & [2]\\
          IUDF & $    70145 $ &   Labb\'e & $2010-2011$ & $  103.0 $ & $  4 $ & $  215.6 $ & $   8464 $ &              S18.18.0/S19.0.0    & [4]\\
          ERS & $    70204 $ &     Fazio & $2011$ & $   95.6 $ & $  2 $ & $  163.0 $ & $   6356 $ &                       S18.18.0   & \\
      S-CANDELS & $    80217 $ &     Fazio & $2011-2012$ &$   25.3 $ & $  4 $ & $  101.2 $ & $   3944 $ &              S19.0.0 /S19.1.0    & [5]\\
        IGOODS & $    10076 $ &     Oesch & $2014$ & $   45.5 $ & $  2 $ & $   65.6 $ & $   2576 $ &                       S19.1.0    & [4]\\
        GREATS & $    11134 $ &   Labb\'e & $2015-2016$ &$   71.8 $ & $  4 $ & $  234.3 $ & $   9200 $ &              S19.1.0 /S19.2.0    & [6]\\
       \hline
        & \multicolumn{2}{r}{Totals\tablenotemark{i}:} & & $315.8$\tablenotemark{$\ddagger$} & $79$ & $1201.3$ & $43278$ & & \\       
\enddata
\tablecomments{Programs PID 81 and PID 20708 were omitted because  they only contribute $0.14$\,hr and $0.5$\,hr depth per pixel and per band, respectively, over the central parts of the GOODS-S region.}
\tablenotetext{a}{Spitzer Program ID}
\tablenotetext{b}{Principal investigator name}
\tablenotetext{c}{Time frame over which observations were carried out.}
\tablenotetext{d}{Maximum coverage depth in hours provided by the program across the region of the field considered in this work}
\tablenotetext{e}{Number of independent pointings}
\tablenotetext{f}{Total observing time (in hours) per band over the field. For cryogenic programs this quantity refers to the cumulative frame time in each of the $3.6$\,$\mu$m, $4.5$\,$\mu$m, $5.8$\,$\mu$m and $8.0$\,$\mu$m bands, while for warm-mission observations it refers to the cumulative frame time in each of the $3.6$\,$\mu$m and $4.5$\,$\mu$m bands.}
\tablenotetext{g}{Total number of basic calibrated data (bcd) frames per band overlapping with the field considered in this work. For cryogenic programs this quantity refers to each of the  $3.6$\,$\mu$m, $4.5$\,$\mu$m, $5.8$\,$\mu$m and $8.0$\,$\mu$m bands, while for warm-mission observations it refers to the $3.6$\,$\mu$m and $4.5$\,$\mu$m bands.}
\tablenotetext{h}{References. Numbers correspond to: [1] \citet{dickinson2003}; [2] \citet{ashby2013a}; [3] \citet{labbe2013}; [4] \citet{labbe2015}; [5] \citet{ashby2015}; [6] This work.}
\tablenotetext{i}{These totals are obtained combining observations from all programs in the $3.6$\,$\mu$m band.}
\tablenotetext{\dagger}{Program executed during the cryogenic part of the mission, providing coverage in the $5.8$ and $8.0$\,$\mu$m bands as well.}
\tablenotetext{\ddagger}{From the coverage map in the $3.6$\,$\mu$m band.}
\end{deluxetable*}

\begin{figure*}
\includegraphics[width=18cm]{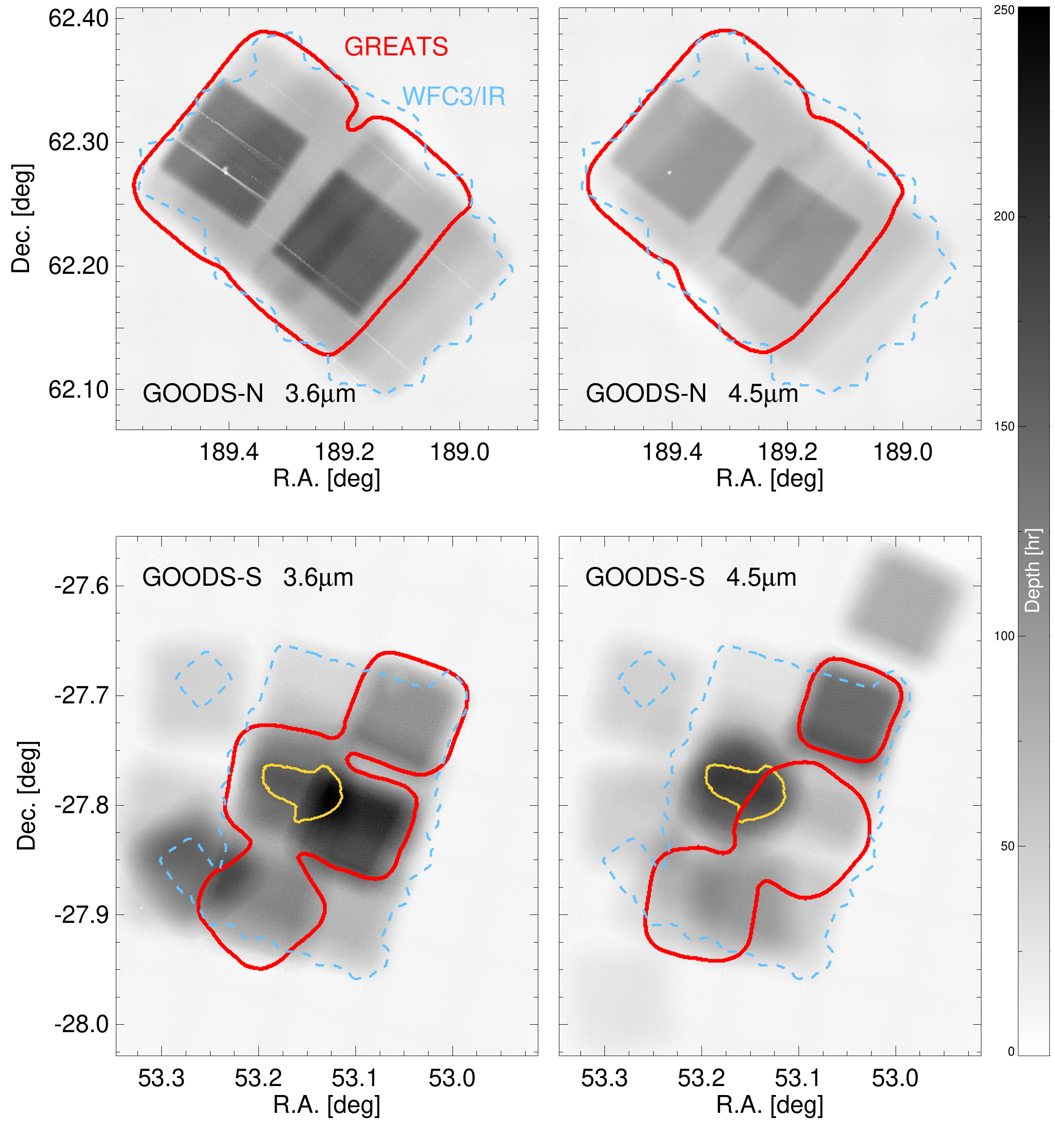}
\caption{The inverted grey-scale images present the coverage maps, in units of hours, in the $3.6$\,$\mu$m (left) and $4.5$\,$\mu$m band (right) for the GOODS-N and GOODS-S fields (top and bottom row, respectively) from all IRAC programs prior to  GREATS. The red  contours mark the coverage provided by the GREATS program, while the light blue dashed contour corresponds to the footprint of the stacked CANDELS/3D-HST WFC3/IR F125W-, F140W- and F160W-band mosaics.  The yellow contours displayed for GOODS-S correspond to the small ($\lesssim10$\,arcmin$^2$) region with a depth $\ge150$\,hr in both bands prior to GREATS. The same grey scale is adopted for all the images to illustrate the depth as indicated by the vertical bar on the right, in units of hours. \label{fig:gns_layout}}
\end{figure*}

During the last $\sim20$ years, the two fields of the Great Observatories Origins Deep Survey initiative (GOODS-N and GOODS-S - \citealt{giavalisco2004}) have accumulated an impressive array of observations ranging from the X-rays to the radio. In particular, the improvements in sensitivity and resolution provided by the \textit{Hubble Space Telescope} (\textit{HST}) Wide Field Camera 3 (WFC3 - \citealt{kimble2008}) in 2009 fostered the acquisition of exquisitely deep optical and near-infrared (NIR) data over these fields through programs such as the Cosmic Assembly Near-infrared Deep Extragalactic Legacy Survey (CANDELS - \citealt{grogin2011, koekemoer2011}), 3D-HST (\citealt{vanDokkum2011,brammer2012}) and the Hubble Ultra/Extreme Deep Field campaigns (UDF09/UDF12/XDF - \citealt{oesch2010,bouwens2010,ellis2013, illingworth2013}). 

These observations have enabled the identification of $\sim400$ plausible galaxies at $z>8$ (e.g., \citealt{bouwens2015, finkelstein2015a}), probing epochs as early as $z\sim10-12$ (e.g., \citealt{oesch2010, oesch2013, oesch2016,  oesch2018, bouwens2011, bouwens2013, bouwens2014, bouwens2019a, ellis2013} - but see also  \citealt{coe2013, zitrin2014, calvi2016, mcleod2016, salmon2018, livermore2018, morishita2018,salmon2020} for similar searches over different fields).

At $z\gtrsim4$ \textit{HST}/WFC3 observations only probe the rest-frame ultraviolet (UV). The \textit{Spitzer} InfraRed Array Camera (IRAC - \citealt{fazio2004}) provided a crucial extension into the rest-frame optical at $4\lesssim z\lesssim 10$, key for studying the stellar mass assembly (e.g. \citealt{duncan2014, grazian2015, song2016, stefanon2017b, davidzon2018}), and nebular line emission (e.g., \citealt{labbe2013,smit2014,faisst2016,debarros2019}) at early cosmic epochs. 

Obtaining coverage of the GOODS fields with \textit{Spitzer} had already been recognized as a priority for the first year of its operations  (\citealt{giavalisco2004}). Indeed, during its 5 years 9 months of cryogenic mission and the following $10$ years $8$ months of warm observations (with an end-of-mission on January 30th, 2020), the GOODS fields have been targeted with IRAC  by a dozen major programs (see Table \ref{tab:programs}), totaling $\sim3$ months of observations per IRAC band.

In this paper we present full-depth mosaics in the IRAC $3.6$, $4.5$, $5.8$ and $8.0$\,$\mu$m bands which combine all relevant observations over the GOODS-N and GOODS-S fields. Specifically, in this release we include new observations at $3.6$ and $4.5$\,$\mu$m  from the \textit{GOODS Re-ionization Era wide-Area Treasury from Spitzer} program (GREATS - PI: I. Labb\'e). All the observations were processed using the latest calibrations and combined together into final mosaics using the same procedure as we earlier pioneered in \citet{labbe2015}. These procedures resulted in a consistent set of data products similar to those produced in that earlier study (see also \citealt{damen2011} for further details). Thanks to its superb depth, this dataset constitutes a natural extension at $3-8$\,$\mu$m of the Hubble Legacy Field initiative (HLF - \citealt{illingworth2016, whitaker2019, illingworth2020}). We also release the photometry in the four IRAC bands, obtained after removing the contamination from neighbours, for $9192$ candidate Lyman-Break galaxies at $z\sim3.5-10$ identified by \citet{bouwens2015} over the GOODS-N and GOODS-S fields.

This paper is organized as follows. In Section \ref{sect:data} we describe the strategy of the observations. Section \ref{sect:data_reduction} summarizes the main steps involved in the reduction of the observations, mosaic and PSF creation. In Section \ref{sect:results} we present the main features of the mosaics, in Section \ref{sect:phot} we describe the procedure we followed to extract the photometry from the GREATS mosaics, while in Section \ref{sect:science} we briefly highlight several science cases motivating the GREATS program. In Section \ref{sect:dr} we specify the data products made available to the community, with the conclusions in Section \ref{sect:conclusions}.

\begin{deluxetable*}{cccccc}
\tablecaption{Summary of GREATS AORs \label{tab:aor}}
\tablehead{\colhead{Field} & \colhead{AOR\tablenotemark{a}} & \colhead{R.A.\tablenotemark{b}} & \colhead{Dec.\tablenotemark{b}} & \colhead{Pos. Angle\tablenotemark{c}} & \colhead{MJD\tablenotemark{d}} \\
\colhead{Name} & &\colhead{[degrees]}  & \colhead{[degrees]} & \colhead{[degrees]} & \colhead{[days]} 
}
\startdata
GOODS-N & $   54396672$ & $ 189.2613220$ & $  62.1838341$ & $ 145.60$ & $  57242.3203125$ \\
&	$   54396416$ & $ 189.2128754$ & $  62.1937408$ & $ 145.46$ & $  57242.4375000$ \\
&	$   54396160$ & $ 189.2434082$ & $  62.1968384$ & $ 145.39$ & $  57242.5507812$ \\
&	$   54395904$ & $ 189.2602844$ & $  62.1841393$ & $ 145.31$ & $  57242.6679688$ \\
& & & & & \\
GOODS-S & $   54319616$ & $  53.1282692$ & $ -27.8142986$ & $  69.03$ & $  57140.3789062$ \\
&	$   54318848$ & $  53.0915413$ & $ -27.8234711$ & $  69.13$ & $  57140.4921875$ \\
&	$   54316032$ & $  53.1064072$ & $ -27.8310547$ & $  69.20$ & $  57140.6093750$ \\
&	$   54317056$ & $  53.1110954$ & $ -27.8082600$ & $  69.56$ & $  57141.1210938$ \\
\enddata
\tablecomments{This Table only presents data for the first four AORs in each field. The full list of AORs is available online. The execution time of each AOR is $\sim2.75$h, with a frame time of $100$s and exposure time of $93.6$s per frame. Each GOODS-N AOR covers an area of $\sim50$arcmin$^2$ per band, while each GOODS-S AOR covers $\sim25$arcmin$^2$ per band. }
\tablenotetext{a}{Astronomical Observation Request (AOR) unique identifier.}
\tablenotetext{b}{Right Ascension and Declination. Positions correspond to the 3.6\,$\mu$m array center of the first frame taken for the given AOR.}
\tablenotetext{c}{Position angle of the array, in degrees East of North.}
\tablenotetext{d}{Modified Julian Day, MJD$\equiv$JD$-2400000.5$, in UTC at start of the first observation in the AOR.}
\end{deluxetable*}

\section{Data}
\label{sect:data}

\begin{figure}
\includegraphics[width=8cm]{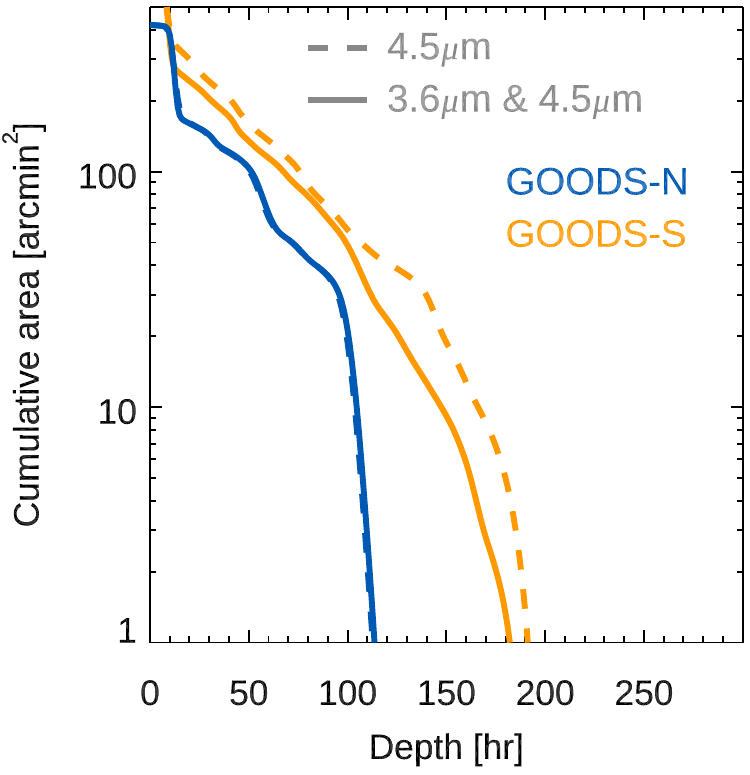}
\caption{Cumulative area as a function of coverage depth (in hours) for the $4.5$\,$\mu$m band (which has the shallowest coverage of the two bands) for the mosaic combining all observations prior to adding in the GREATS compared to the cumulative area where $3.6$ and $4.5$\,$\mu$m bands simultaneously benefit from the same minimum coverage depth (dashed and solid curves, respectively), for the GOODS-N and GOODS-S fields (blue and orange curve, respectively). \label{fig:match}}
\end{figure}

The GOODS fields are centered at $\alpha=$12$^h$36$^m$55$^s$, $\delta=$+62\degree14$'$15$''$ (GOODS-N) and $\alpha=$03$^h$32$^m$30$^s$, $\delta =$-27\degree48$'$20$''$ (GOODS-S), respectively, and benefit from extensive IRAC coverage. Specifically, our mosaics combine essentially all observations from past programs to new observations at $3.6$\,$\mu$m and $4.5$\,$\mu$m from the GREATS program (PI: I. Labb\'e).  The main properties of the programs included in our analysis are summarized in Table \ref{tab:programs}.

\subsection{Archival coverage}
 The GOODS-N field was observed for $558.6$\,hr in the $3.6$ and $4.5$\,$\mu$m each and for $278.3$\,hr in each of the $5.8$ and $8.0$\,$\mu$m bands, while GOODS-S was observed for $967.0$\,hr in each of the $3.6$ and $4.5$\,$\mu$m bands and $209.8$\,hr at $5.8$ and $8.0$\,$\mu$m, for a total of $4027.4$h. Observations in the $5.8$ and $8.0$\,$\mu$m bands were feasible only during the cryogenic part of the mission, making the  accumulation of data in these channels significantly shorter than is available in the  $3.6$ and $4.5$\,$\mu$m bands.

\subsection{GREATS observing strategy and coverage}

Our Cycle 11 GREATS program added significantly to the GOODS archival datasets, primarily by adding the crucial deep data needed for a number of wide-ranging science investigations. GREATS contributed $606.9$\,hr of additional coverage in each of the $3.6$ and $4.5$\,$\mu$m bands, bringing the cumulative coverage in each of these bands to $\sim2132.5$\,hr, and in the four bands to $5241.2$\,hr. 

Consistent photometric depth across multiple wavelength channels is key for obtaining detailed probes of the spectral energy distribution (SED) of high redshift galaxies (\citealt{labbe2015}). Unfortunately, the different observational programs over the GOODS fields led to rather heterogeneous datasets, resulting in very position-dependent IRAC depths. GOODS-S suffered more clearly from this issue: ultradeep ($\gtrsim150$\,hr) coverage existed in both the $3.6$ and $4.5$\,$\mu$m bands; yet, only a tiny $\sim10$\,arcmin$^2$ area was observed at this depth in both bands, limiting the scientific value of the dataset. This can qualitatively be seen in Figure~\ref{fig:gns_layout}, and, more quantitatively, in Figure \ref{fig:match}.

The main aim of the GREATS program was to significantly extend the ultradeep $\sim150$\,hr coverage to $\gtrsim150$\,arcmin$^2$ across both GOODS fields, while at the same time improving the overall homogeneity of the $3.6$ and $4.5$\,$\mu$m band coverage. The layout of the observations was carefully chosen to complement and expand the already existing data, taking advantage of the change in position angle of the arrays as  \textit{Spitzer} travelled along its orbit, maximizing the survey efficiency. The full set of Astronomical Observation Requests (AORs) from the GREATS program are listed in Table \ref{tab:aor}. Observations over the GOODS-N field were split into two pointings executed  $\sim180$ days apart, with each pointing covering the rectangular area of two contiguous arrays.  To gain more uniform depth in the GOODS-S field required us to organize the observations into four pointings, each one probing the region corresponding to one array.  All AORs were executed with a \textit{medium cycling} dithering pattern. The resulting additional coverage is shown in Figure \ref{fig:gns_layout}.

\section{Data reduction and mosaic creation}
\label{sect:data_reduction}

To limit systematics and to generate uniformly-processed mosaics, we downloaded from the Spitzer Heritage Archive all the relevant observations from all the programs overlapping the CANDELS GOODS footprint. The full dataset consists of 845 AORs (392 + 453, for GOODS-N and GOODS-S, respectively), for a total of 168234 individual frames (72494 in GOODS-N and 95740 in GOODS-S, respectively).

The data reduction started with the most recent corrected Basic Calibrated Data (cBCD) generated by the \textit{Spitzer} Science Center (SSC) calibration pipeline. This subtracts the dark frames, homogenizes the pixel response (detector linearization and flat fielding), corrects  for known artifacts (column pull up/down, muxbleed, first frame effect), and provides per-frame uncertainty estimates, bad pixel mask, and cosmic ray rejection masks.

We post-processed the cBCD frames following the same custom pipeline used by \citet{labbe2015}. This pipeline improves on these initial corrections and masks (see Sec. 3.1 of \citealt{labbe2015}), combines the frames and generates the final mosaics. In the following Section we summarize the pipeline's main steps, referring the reader to \citet{damen2011} and \citet{labbe2015} for further details.

\subsection{IRAC reduction process}

\begin{figure}
\includegraphics[width=9cm]{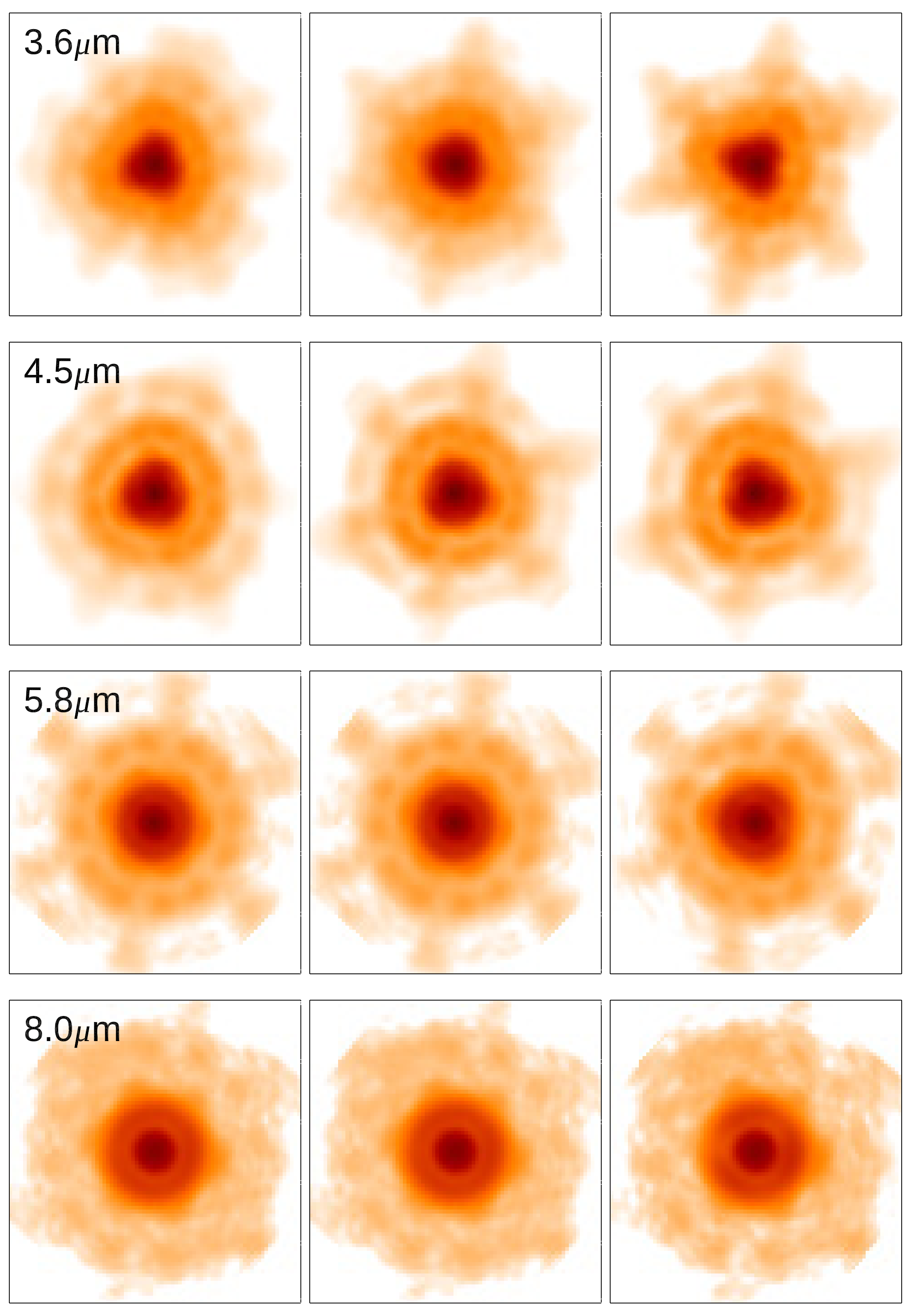}
\caption{Examples of point-spread functions (displayed with a logarithmic intensity map) at different positions across the GOODS-N field.  Each row refers to a specific IRAC band, as indicated by the label at the top-left corner of the left-most panel. In each row, the three panels, $\sim24''$-wide, show PSFs in the GOODS-N field at same declination, but spaced in R.A. by $\sim15'$.  They  highlight the rapid spatial variation of the PSFs across the IRAC field, particularly evident for the $3.6$ and $4.5$\,$\mu$m bands. A similar variation is seen in GOODS-S. Such dramatic variations clearly complicate the analysis and the photometry. \label{fig:psf}}
\end{figure}

In general, cBCDs from different programs were generated using different SSC pipeline versions (see Table \ref{tab:programs}). The main differences consist of astrometry, image distortion refinements, and artifact correction. These issues were all handled directly in our own reduction pipeline, and hence these updates by the SSC have no effect on our end products.  We note here that the flux density calibration has not changed significantly since S18.8, and therefore our mosaics use the latest flux calibrations consistent with S19.2.

The reduction with our custom pipeline is based on a two-pass procedure, where each AOR was reduced independently. The first pass included the following steps: an initial removal of background  and bias structure from each frame estimated from the median of all the frames in the AOR; correction of column pull-up and pull-down introduced by bright stars or cosmic rays subtracting a median above and below the affected pixels after excluding any sources; persistence masking  and muxbleed correction rejecting all highly exposed pixels in the subsequent $4$ frames.

\begin{figure*}
\includegraphics[width=18cm]{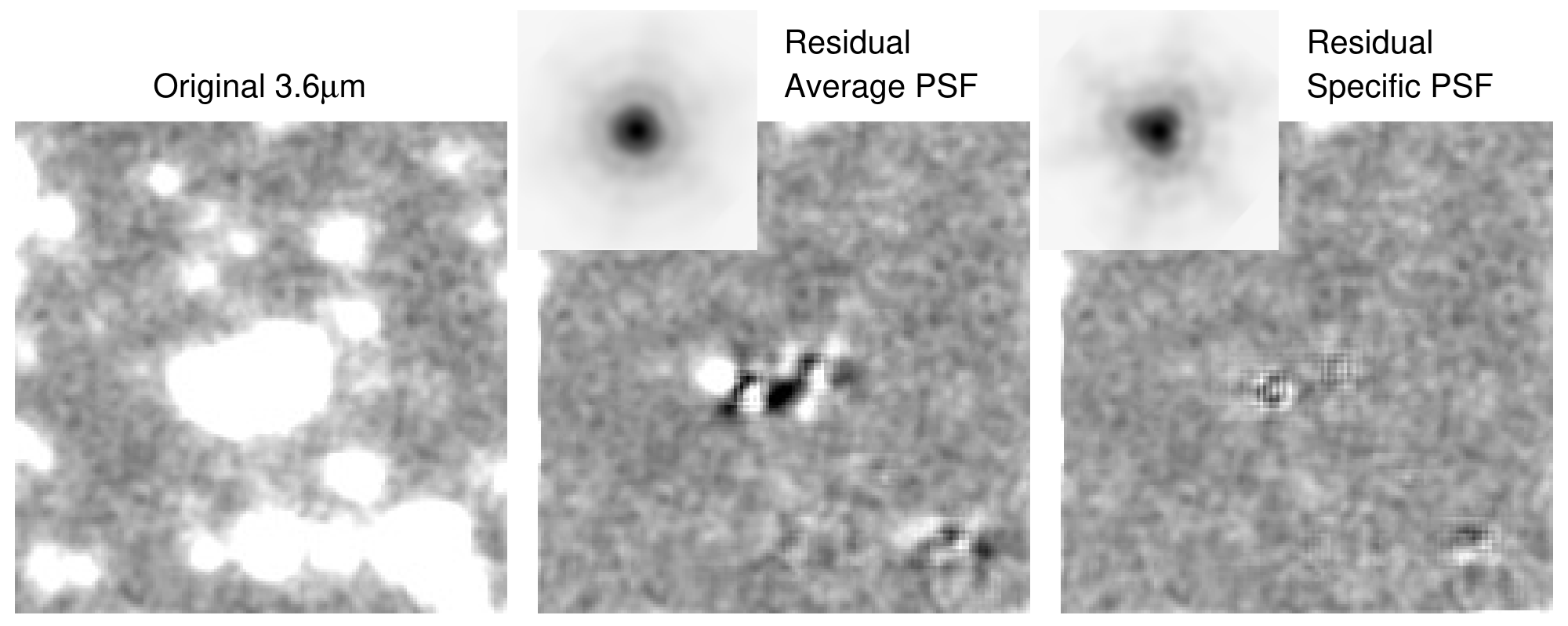}
\caption{Impact of the PSF construction method on prior-based photometry. The panel on the left shows an image stamp at a random location in the GREATS GOODS-N  $3.6$\,$\mu$m science mosaic. The middle and right panels present two different residuals after subtracting from the science image a model built convolving the 3D-HST detection image \citep{skelton2014} with a convolution kernel based on the PSF in the $3.6$\,$\mu$m band displayed in the inset at the corresponding top-left corner (using a logarithmic stretch and inverted grey scale). The central panel shows the residual from subtracting the average PSF over the full GOODS-N field (comparable to the PSF built by stacking isolated point sources across the field), while the panel on the right shows the residual using our position-dependent PSF, as pioneered by \citet{labbe2015}. All image stamps are $27\farcs5$ per side. The position-dependent PSF minimizes systematics in the photometry, and clearly produces superior photometric results. \label{fig:resid}}
\end{figure*}

 The second pass included  cosmic ray rejection, astrometric calibration and an accurate large-scale background removal. Cosmic ray hits were cleaned through iterative sigma clipping. The astrometry was corrected applying a rigid shift in both R.A. and Dec. estimated from sources in common with the deep maps of CANDELS/3D-HST \citep{skelton2014}. The absolute astrometric reference of CANDELS/GOODS-N was registered to the Sloan Digital Sky Survey, the Two Micron All Sky Survey (2MASS), and the deep Very Large Array (VLA) $20$\,cm survey (\citealt{morrison2010}), while that for CANDELS/GOODS-S was anchored to the $R$-band mosaic from the ESO Imaging Survey (EIS - \citealt{arnouts2001}), registered to the Guide Star Catalog II (GSC-II - \citealt{lasker2008} - \citealt{koekemoer2011}). The background level was first estimated as the median of the frames in each AOR masking sources and outlier pixels, and refined by iteratively clipping pixels belonging to objects and subtracting the mode of the background pixels. The frames were then drizzled \citep{fruchter2002} using a pixfrac = $0.4$ on a common reference grid defined by the CANDELS tangent point and a fine $0\farcs3$ pixel$^{-1}$ scale, to allow for easy re-binning onto commonly adopted pixel scales. The individual drizzled AORs were combined into the final mosaic after weighting each pixel according to its depth.

The output pixels in the final drizzled image are not independent of each other, causing the pixel-to-pixel noise in the output image to be correlated. The correlation implies that direct estimates of the pixel-to-pixel noise in the drizzled output image underestimates the noise on large scales. For the drizzle parameters used here, an approximate correction from the single-pixel noise to the noise at large scales can be derived following \citet{casertano2000}. At an output pixel scale of $0\farcs3$ (scale $= 0.25$ relative to the $1\farcs2$ input pixel) each (pixfrac$=0.4$) drizzled input pixel contributes to $1.6$ output pixels. In this case, the noise at large scales is a factor $2.0$ higher than estimated from the pixel-to-pixel rms (see \citealt{casertano2000} Appendix A6 for pixfrac $>$ scale). An output image at a scale of $0\farcs6$ can easily be produced from the $0\farcs3$ scale images by simply block summing the science images and weight maps by a factor $2$ in each dimension. The resulting image at $0\farcs6$ scale has drizzle pixfrac $= 0.4$ and scale $= 0.5$, so each drizzled pixel contributes to $< 1$ output pixel. In this case, the noise over large areas is a factor $1.36$ higher than estimated from the pixel-to-pixel noise.

\begin{figure*}
\includegraphics[width=18cm]{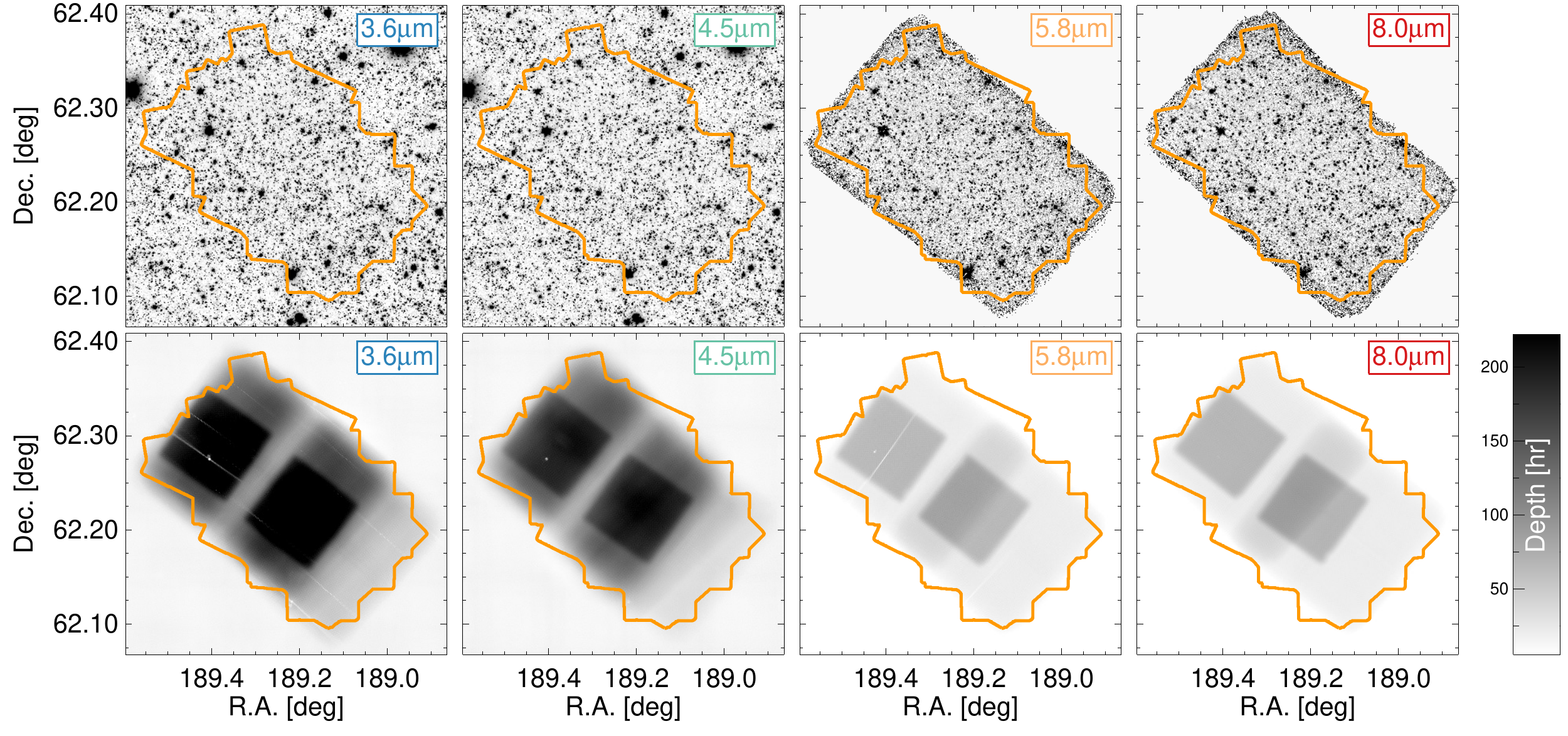}
\caption{Full GREATS mosaics in the IRAC $3.6$, $4.5$, $5.8$ and $8.0$\,$\mu$m  bands (left to right, respectively) for the CANDELS/GOODS-N field. The top row presents the science mosaics in inverted grey scale, while the bottom row shows the coverage maps, with the effective depth indicated by the vertical bar on the right. The orange contour in each panel corresponds to the region covered by the 3D-HST detection mosaic \citep{skelton2014}, combining the data in the \textit{HST}/WFC3 F125W, F140W and F160W bands.  The coverage in the $3.6$ and $4.5$\,$\mu$m beyond the CANDELS boundary is provided by the SEDS program.\label{fig:gn_greats}}
\end{figure*}

\subsection{Point-spread function creation}
\label{sect:psf}
The instrumental point-spread function (PSF) in the IRAC bands, particularly in the $3.6$ and $4.5$\,$\mu$m bands, shows a peculiar, approximately triangular shape (see. e.g., the \textit{Spitzer Space Telescope} Observer Manual - Sect. 6.1.2.2).  AOR observations spread over periods of time of the order of months cause observations of the same patch of the sky to have position angles differing by tens of degrees. These angular offsets result from a change in the spacecraft roll angle of approximately $1$\,deg per day.  The combination of the instrumental PSF at different position angles results in complex light profiles for the final PSFs, which  can also change rapidly over small spatial scales. Accurate reconstruction of the PSF, then, becomes an essential step to obtain robust IRAC photometry using either PSF- or prior-based fitting techniques. However, identifying a suitable number of high-S/N isolated point sources at different locations across the field is generally a challenging task in deep extragalactic fields due to source crowding.

\begin{figure*}
\includegraphics[width=18cm]{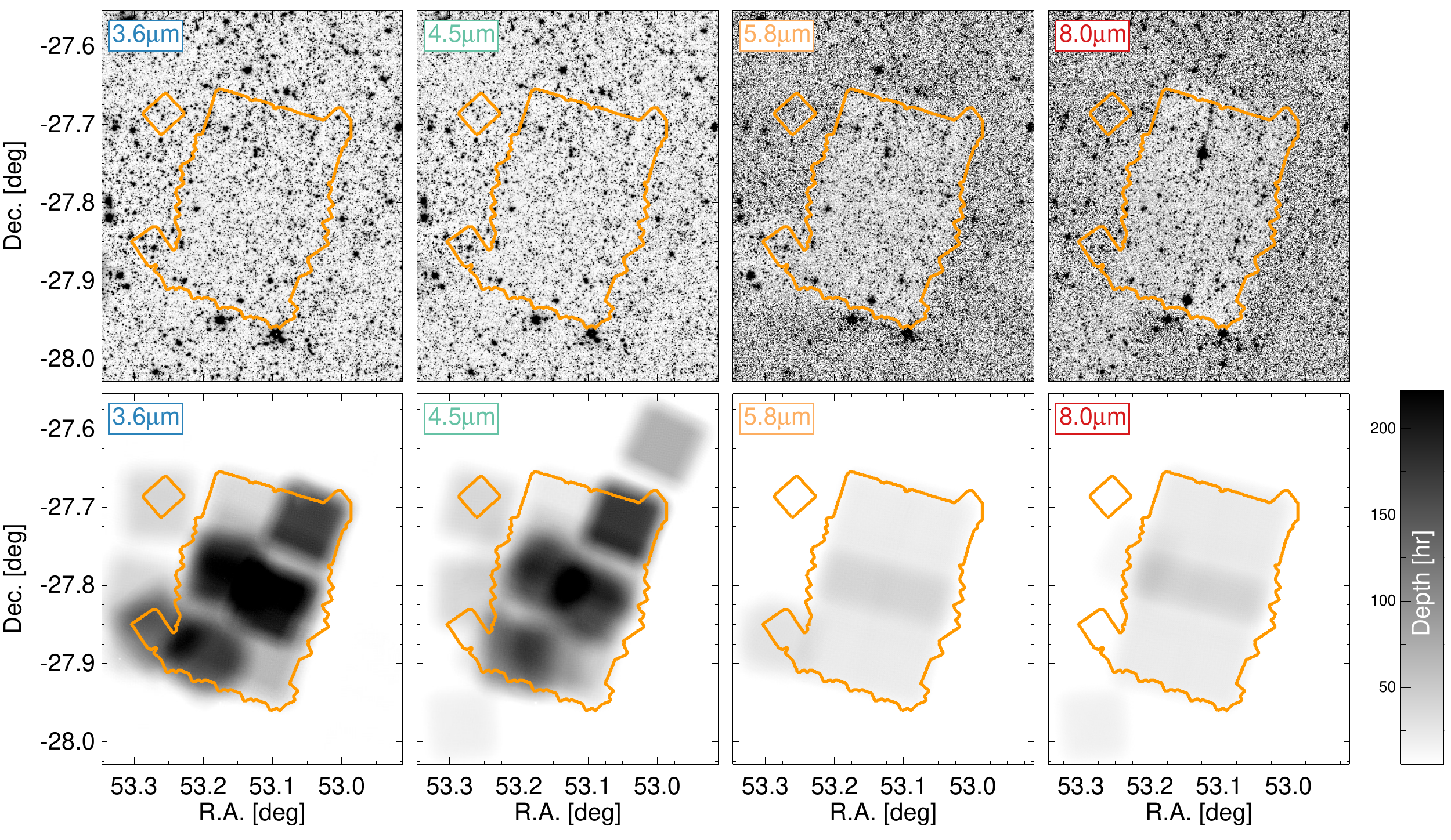}
\caption{Full GREATS mosaics in the IRAC $3.6$, $4.5$, $5.8$ and $8.0$\,$\mu$m  bands (left to right, respectively) for the CANDELS/GOODS-S field. The top row presents the science mosaics in inverted grey scale, while the bottom row shows the coverage maps, with the effective depth indicated by the vertical bar on the right. The orange contour in each panel corresponds to the region covered by the 3D-HST detection mosaic \citep{skelton2014}, combining the data in the \textit{HST}/WFC3 F125W, F140W and F160W bands. The coverage in the $3.6$ and $4.5$\,$\mu$m beyond the CANDELS boundary is provided by the SEDS program. \label{fig:gs_greats}}
\end{figure*}

To overcome this problem, we leveraged the remarkable instrumental stability of IRAC over its life cycle. We created extremely high S/N empirical PSFs in the four IRAC bands stacking several hundred observations of bright, unsaturated point sources (see \citealt{labbe2015} for full details), and we adopted these as a template. These templates extend to a radius of $12\farcs0$ and share the pixel scale of the science mosaics ($0\farcs3$/pixel - see Section \ref{sect:results}); location-dependent PSFs were then generated combining the template by rotating and stacking according to the position angles and coverage depth of each AOR stored in a fine grid (steps of $12\farcs0$) of locations across each mosaic. 

Our PSF reconstruction procedure takes advantage of the approximate invariance of the effective PSF of the IRAC array across its field of view (FoV). Comparison between $3\farcs0$-diameter aperture photometry from the warm mission on the $100\times$-oversampled IRAC Point Response Function (PRF - IRAC Instrument Handbook) at the edges of the IRAC FoV to that from the PRF at the center of the array resulted in systematic differences of $\lesssim3\%$. Most importantly, the GREATS mosaics combine, at each point on the sky, observations from different programs. These sampled the same patch of sky at different portions of the IRAC array, averaging over the effective PSF of all contributing exposures. The end result is a finely sampled PSF at each location in the GOODS fields.

In Figure \ref{fig:psf} we present examples of the PSF variation in the $3.6$\,$\mu$m band. These examples highlight the dramatic changes in the shape of the PSF even across small regions of the mosaics. They indicate that the adoption of  non-optimal PSFs in prior-based photometry may introduce systematic effects. 
We further illustrate this in Figure \ref{fig:resid}. The strong variation of the PSF profile across the mosaics makes the adoption of the average PSF much less effective at correctly reproducing the observed light profiles of the objects. The resulting residuals are substantial and can introduce systematic errors in the flux density estimates. The adoption of the optimal PSF results in marginal residuals (see also e.g., Figure 4 of \citealt{merlin2016}). As an aide to future photometric studies, we release together with our mosaics, the template PSF with the AOR mapping data and a Python script to reconstruct the PSF at arbitrary locations across the mosaics.

\section{Results}
\label{sect:results}

\subsection{Mosaic properties}

\begin{figure*}
\includegraphics[width=17cm]{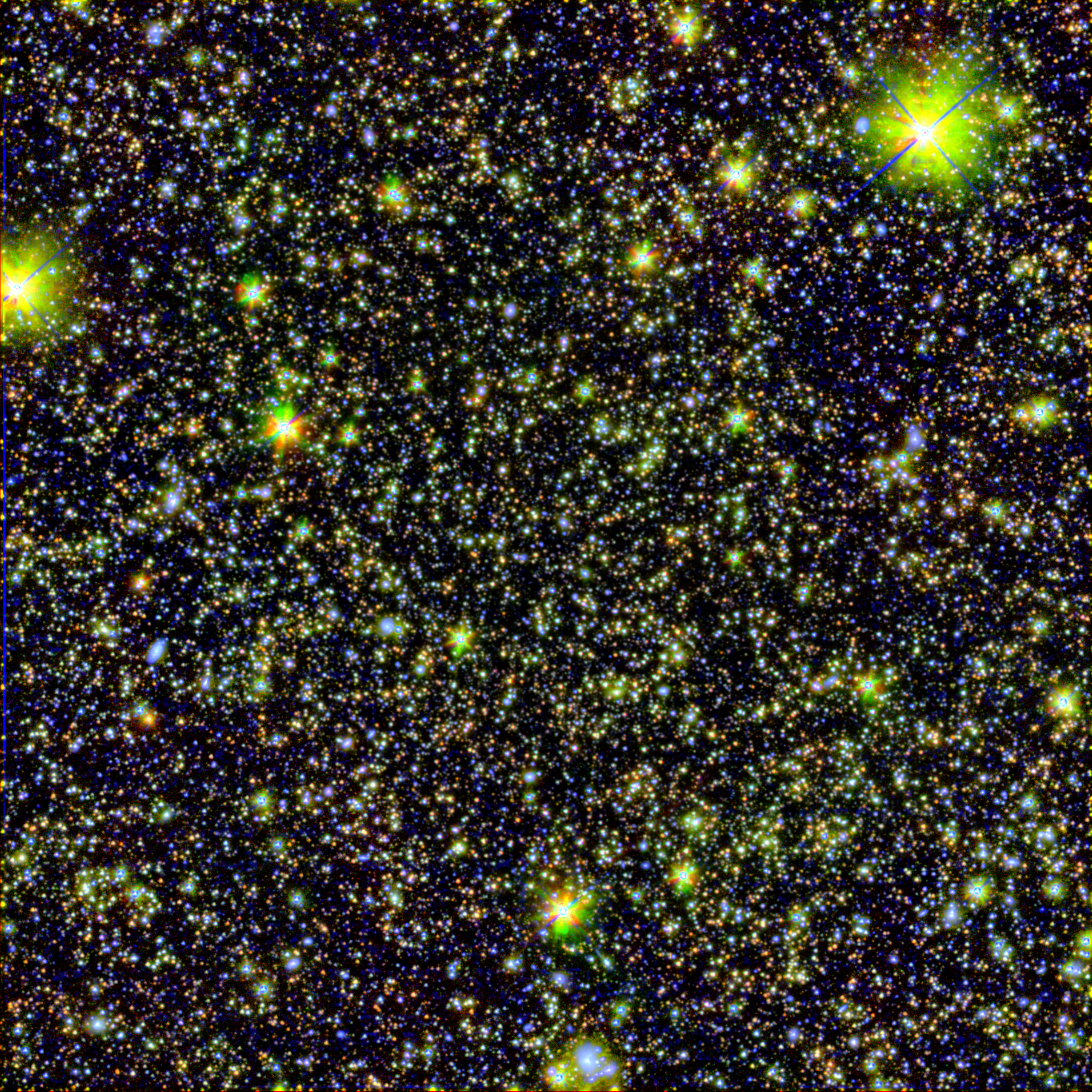}
\caption{Color composite image of the GOODS-N field ($\sim 20'\times 20'$) obtained combining the new \textit{Spitzer}/IRAC imaging from the GREATS program (red and green respectively show the IRAC $4.5$ and $3.6$\,$\mu$m emission) with the $K_\mathrm{s}$ data (shown in blue) from the MODS (\citealt{kajisawa2011}) and CFHT $K_\mathrm{s}$ \citep{wang2010} programs.    \label{fig:clr_n}}
\end{figure*}

\begin{figure*}
\includegraphics[width=17cm]{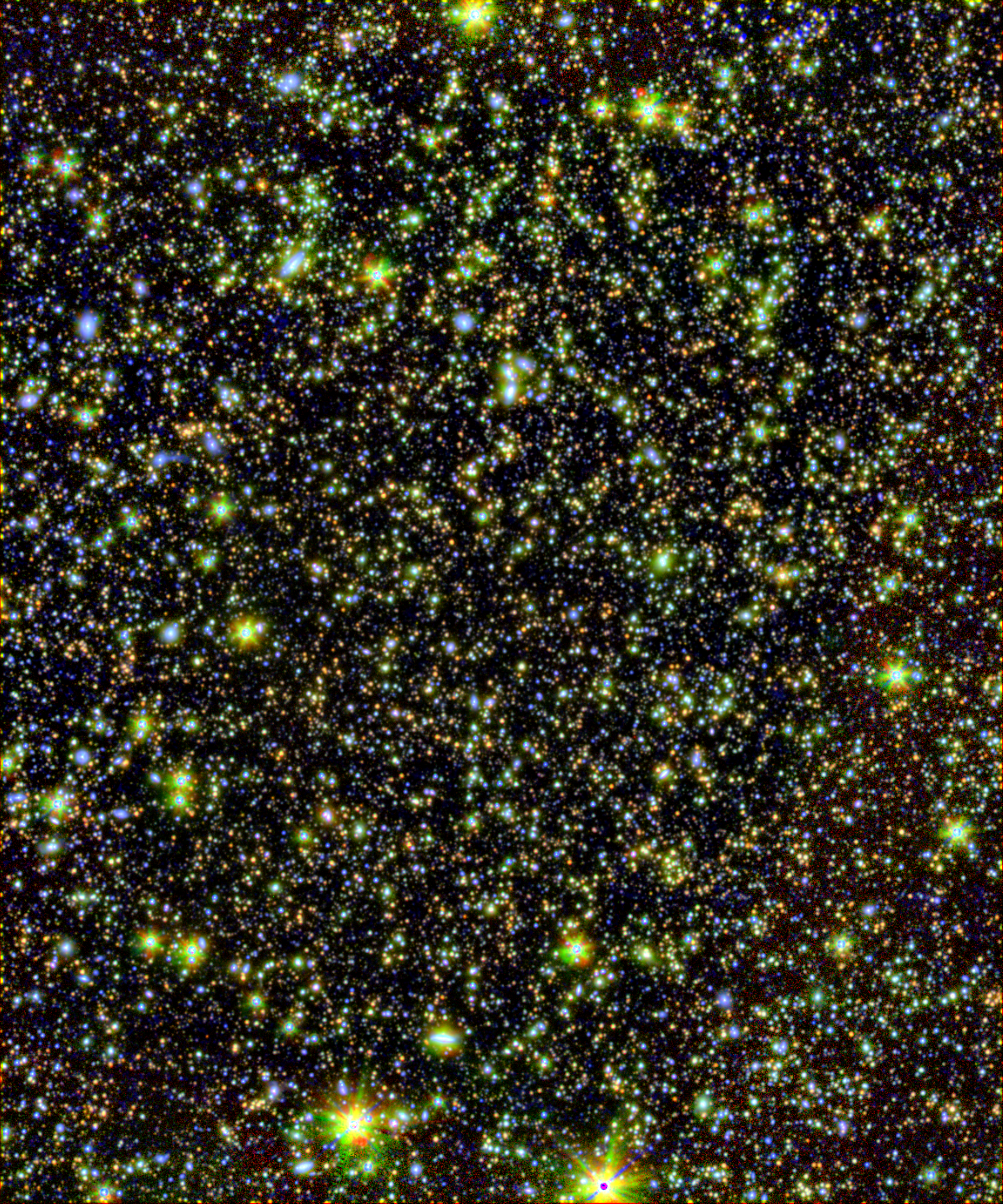}
\caption{Color composite image of the GOODS-S field ($\sim 23'\times 28'$) obtained combining the new \textit{Spitzer}/IRAC imaging from the GREATS program (red and green respectively show the IRAC $4.5$ and $3.6$\,$\mu$m emission) with the $K_\mathrm{s}$ data (shown in blue) from the TENIS (\citealt{hsieh2012}) and HUGS (\citealt{fontana2014}) programs.  \label{fig:clr_s}}
\end{figure*}

The final full-depth IRAC mosaics in the four bands, together with the corresponding coverage maps, are shown in Figure \ref{fig:gn_greats} and \ref{fig:gs_greats} for the GOODS-N and GOODS-S fields, respectively. Color images combining $K_\mathrm{S}$-band, $3.6$ and $4.5$\,$\mu$m are shown in Figures \ref{fig:clr_n} and \ref{fig:clr_s} for GOODS-N and GOODS-S, respectively.

The FWHMs of the PSFs, measured from their radial profile, and their $1$\,$\sigma$ dispersions across the field are $1\farcs52\pm 0\farcs02$ at $3.6$\,$\mu$m for both fields; $1\farcs49 \pm 0\farcs02$ and $1\farcs50 \pm 0\farcs02$ at $4.5$\,$\mu$m (GOODS-N and GOODS-S, respectively); $1\farcs72\pm0\farcs01$ and $1\farcs69\pm0\farcs01$ at $5.8$\,$\mu$m (GOODS-N and GOODS-S, respectively); and  $2\farcs06\pm0\farcs01$ and $2\farcs04\pm0\farcs01$ at $8.0$\,$\mu$m (GOODS-N and GOODS-S, respectively). These value indicate excellent and constant image quality across the two fields; the FWHMs for the two bluer bands are  consistent at $\lesssim2\sigma$ with those of the IUDF program \citep{labbe2015}. 

The photometric calibration was verified using $\sim80$ bright ($m\lesssim 21$ AB) unsaturated point sources spread across each field. We measured their flux densities in large ($6\farcs0$ radius) apertures and find excellent agreement ($<2\%$) with measurements from the S-CANDELS mosaics \citep{ashby2015}. Comparison to the original GOODS mosaics revealed a larger offset, $\sim 8\%$ in the $3.6$\,$\mu$m band and $\lesssim2\%$ in the $4.5$\,$\mu$m band for both fields, consistent with \citet{labbe2015}.  These offset are understood and are due to the different flux calibration of the BCD pipeline used to reduce the original GOODS observations (see \citealt{labbe2015} for details). Based on these comparisons, we estimate the accuracy of our flux density maps to be $2\%$.

\subsection{Astrometric reference}
\label{sect:astrometry}

The registration to the reference CANDELS frame is accurate to $<0\farcs0045$ with a $1$\,$\sigma$ dispersion of $0\farcs077$. Comparison of the astrometry between the CANDELS/3D-HST GOODS mosaics and the second release of the \textit{Gaia} catalogue (\textit{Gaia} DR2 - \citealt{gaia_dr2, lindegren2018}) showed that in each field the astrometric reference suffered from position-dependent distortions with offsets of $\sim0\farcs08$ rms (\citealt{illingworth2020}). Furthermore, the astrometry of GOODS-S is affected by an overall shift of $\sim0\farcs3$ to the North and $\sim0\farcs1$ to the West, consistent with what found by \citet{rujopakarn2016} for the HUDF. Given the offsets to the \textit{Gaia} DR2 astrometric reference are marginal compared with the PSF FWHMs, we ultimately opted not to apply any further correction to the GREATS astrometric solution. The released GREATS mosaics therefore share the same astrometric reference as CANDELS.

We anticipate that the extraction of information from the GREATS mosaics will be performed in one of the following two ways, ultimately depending on whether the sources of interest have a counterpart in a high-resolution image. If a counterpart is available, its IRAC photometry would be best measured with one of the tools developed for prior-based photometry (e.g., \textsc{Tfit} - \citealt{laidler2007}, and descendants) and adopting as prior a mosaic with the same astrometric reference of GREATS (such as those from CANDELS or 3D-HST). Nonetheless, because current tools for prior-based photometry can accommodate small position-dependent shifts ($\sim0\farcs2-0\farcs4$), we expect only marginal systematic effects in photometry even when adopting as prior high-resolution mosaics whose astrometric reference is consistent with that of \textit{Gaia}. The detailed impact of these effects would however depend on the geometry of the individual sources and on the alignment capabilities of the tools adopted for the photometry. Broadly, however, the spatial coordinates of the objects of interest can then be anchored to the \textit{Gaia} astrometric reference  by registering to the coordinates of the object on the high-resolution image. The astrometric accuracy for sources without a high-resolution counterpart will be subject to the $0\farcs08$ scatter discussed above (natively or after applying a rigid offset of (R.A., Dec.) $=(0\farcs1, -0\farcs3)$ for the GOODS-N and GOODS-S mosaics, respectively). However, we would note for any investigations involving direct detection of sources on the IRAC mosaics, that the source positional uncertainties will be dominated by the broad IRAC PSF, likely requiring pre-imaging for follow-up observations where $\sim$few$\times$\,mas accuracy is necessary (e.g., as for \textit{JWST}/NIRSpec - \citealt{bagnasco2007, jdox}).

\begin{figure*}[t!]
\includegraphics[width=18cm]{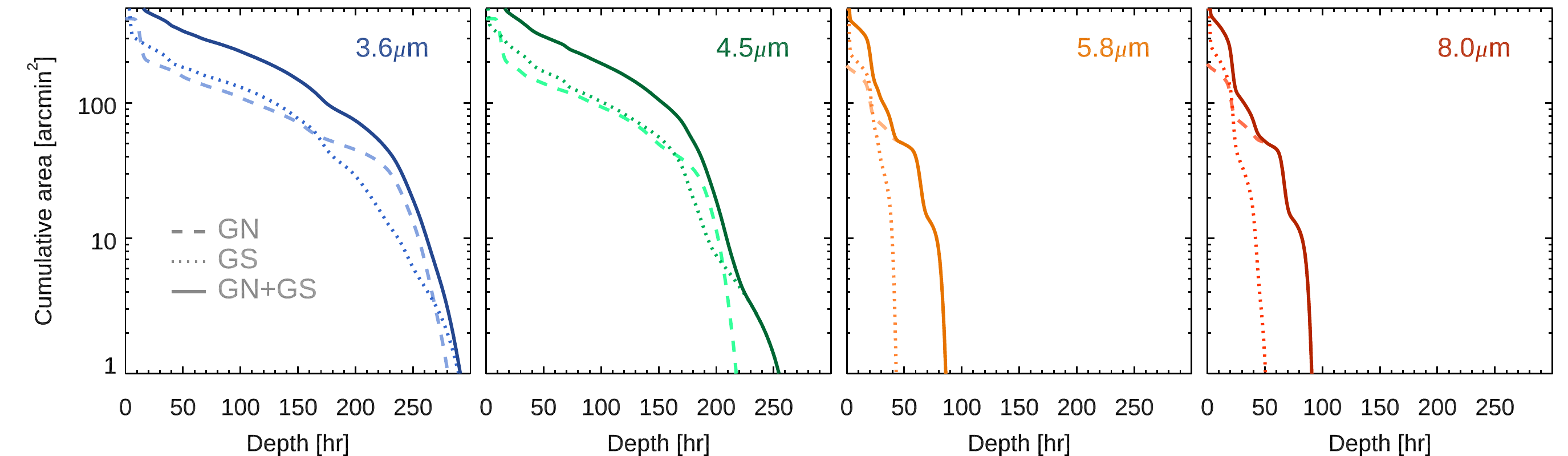}
\caption{Cumulative area as a function of the depth (in hours) for the final GREATS mosaics in the $3.6$, $4.5$, $5.8$ and $8.0$\,$\mu$m bands, respectively, as indicated by the label at the top-right corner. In each panel, the different curves correspond to the area for the CANDELS/GOODS-N, CANDELS/GOODS-S and their combination, according to the legend in the left panel. \label{fig:exptime}}
\end{figure*}

\subsection{Coverage Depth}
\label{sect:exptime}

In Figure \ref{fig:exptime} we present the cumulative area as a function of the coverage depth for the four bands, for the individual fields and when combined together. Overall, the two fields have approximately the same coverage depth distribution in the $3.6$ and $4.5$\,$\mu$m bands, $\sim3-5\times$ deeper than the $5.8$ and $8.0$\,$\mu$m bands. Furthermore, the $5.8$ and $8.0$\,$\mu$m coverage  in GOODS-N is generally $\sim2\times$ deeper than in GOODS-S.

The coverage of the $3.6$\,$\mu$m band reaches a maximum depth of $\sim 250$\,hr over $\sim5-10$\,arcmin$^2$ in each field, while that of the $4.5$\,$\mu$m band reaches $\sim200$\,hr over $\sim10-20$\,arcmin$^2$ area in each field. Remarkably, the mosaics in the $3.6$\,$\mu$m ($4.5$\,$\mu$m) band provide an ultradeep coverage of at least $150$\,hr ($120$\,hr) over a cumulative area of $\sim150$\,arcmin$^2$ ($\sim1/2$ the total area of the CANDELS footprint of the GOODS fields).

\begin{figure}[b!]
\hspace{-0.2cm}\includegraphics[width=8cm]{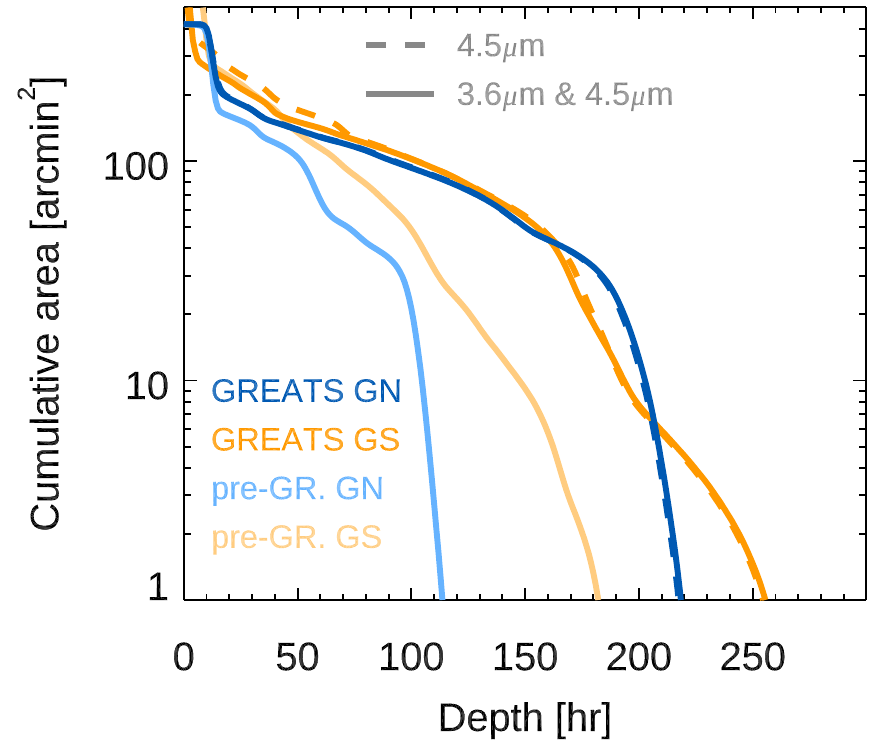}
\caption{Cumulative area vs. per-pixel-depth (in hours) in the $4.5$\,$\mu$m band (dashed curves) and vs. the minimum per-pixel-depth in the $3.6$ and $4.5$\,$\mu$m bands (solid curves), for the GOODS-N and GOODS-S fields (blue and yellow, respectively). Darker colors refer to the GREATS mosaics, while lighter curves correspond to the cumulative areas for the mosaics combining all data available prior to GREATS. These are reproduced here from Figure \ref{fig:match} for comparison.
\label{fig:exptime_comparison}}
\end{figure}

The full-depth coverage maps presented in Figure \ref{fig:gn_greats} and \ref{fig:gs_greats} qualitatively show that for both GOODS fields the coverage depth with the $3.6$\,$\mu$m band is spatially distributed in a very similar way across the mosaic as that in the $4.5$\,$\mu$m band. This is presented more quantitatively in Figure \ref{fig:exptime_comparison}. The substantial overlap between the cumulative area in the $4.5$\,$\mu$m band coverage with that of the minimum coverage between  $3.6$ and $4.5$\,$\mu$m bands indicates that, despite some inhomogeneities in the depth of coverage in each field,  the per-pixel coverage in the $3.6$\,$\mu$m band is at least as deep as that in the $4.5$\,$\mu$m band over most of the GOODS area.  Specifically, for GOODS-S, for $\sim95\%$ of the area with  $>75$\,hr depth coverage, the coverage depth in the $3.6$\,$\mu$m band is at least $90\%$ of that in the $4.5$\,$\mu$m band. This is quite remarkable considering the initial coverage inhomogeneity discussed in Section \ref{sect:data}.

\begin{figure}
\hspace{-0.5cm}\includegraphics[width=9cm]{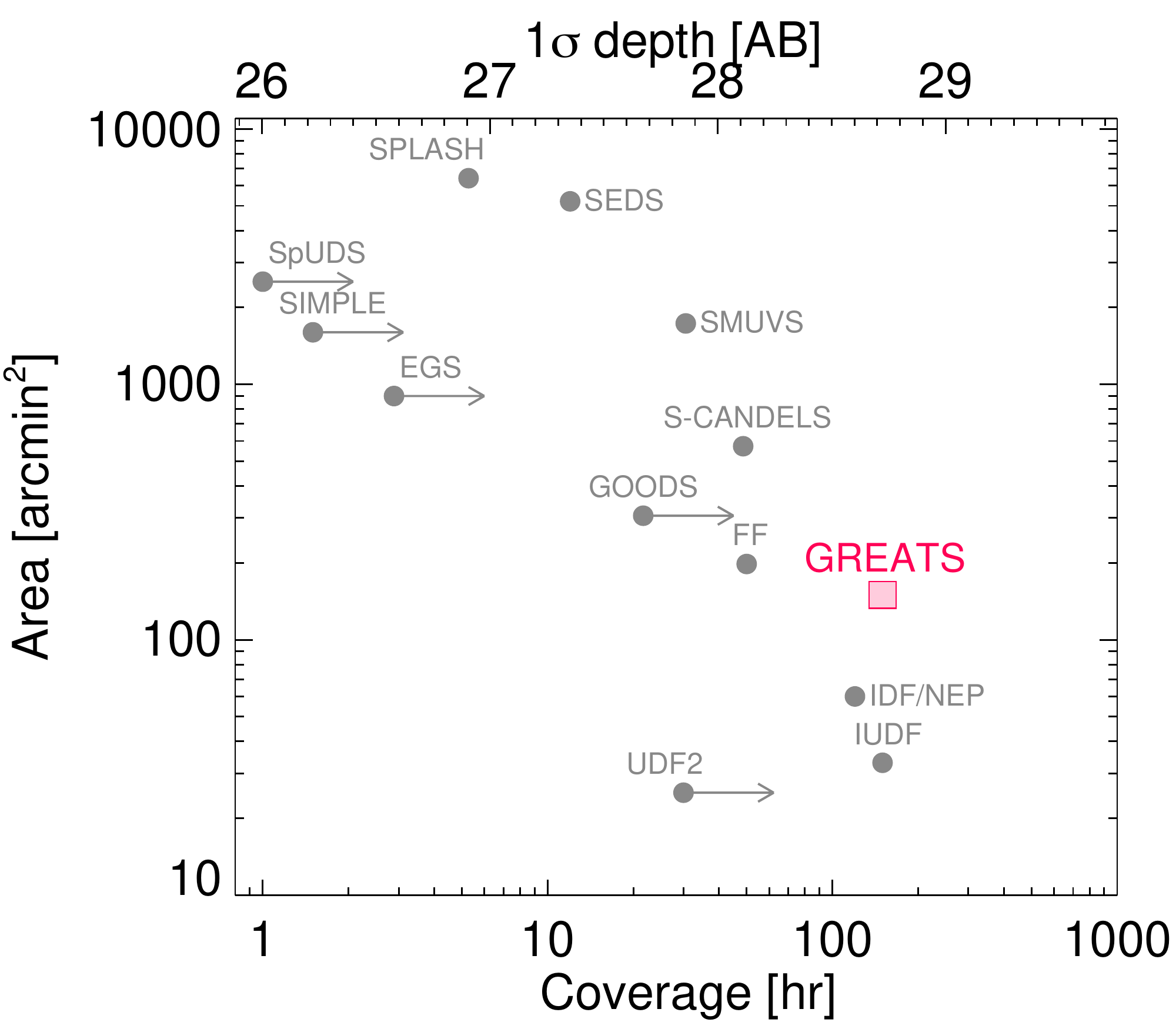}
\caption{The GREATS coverage in the context of the major \textit{Spitzer}/IRAC surveys for high-redshift galaxies. The horizontal axis at the top presents an approximate $1$\,$\sigma$ depth in the $3.6$\,$\mu$m band for point sources  from the SENS-PET calculator (but see the discussion in Section \ref{sect:depth}). Data points refer to the following programs: IUDF \citep{labbe2015}, UDF2 \citep{labbe2013}, SIMPLE \citep{damen2011}, Frontier Fields (FF; PI: Soifer - see e.g., \citealt{shipley2018}), SPLASH \citep{steinhardt2014}, SEDS \citep{ashby2013a}, S-CANDELS \citep{ashby2015}, SMUVS \citep{ashby2018}, SpUDS \citep{caputi2011}, EGS \citep{barmby2008}, GOODS (PI: Dickinson) and IDF/NEP (\citealt{krick2009} - where the data point was extracted from the Data Release 2 mosaic)\textsuperscript{a}. Arrows identify those programs executed during cryogenic cycles, and indicate, by the end point of the arrow, that cryogenic observations were $\sim2\times$ more efficient in the $3.6$\,$\mu$m band than warm mission ones (SENS-PET). Building in part on existing observations, GREATS extends the depth of IUDF to an area $\sim5\times$ larger.} 
\small\textsuperscript{a} \url{http://spider.ipac.caltech.edu/staff/jason/darkfield/styled/index.html}
 \label{fig:surveys}
\end{figure}

Figure \ref{fig:surveys} shows the position of the GREATS full-depth mosaics in the depth-area plane (where we adopt the characteristic depth of $150$\,hr over $150$\,arcmin$^2$ in the $3.6$\,$\mu$m band) and compares it to other significant extragalactic deep surveys executed with \textit{Spitzer}. GREATS provides a depth comparable to that of IUDF but over a $\sim5\times$ larger area.

\subsection{Sensitivity}
\label{sect:depth}

The nominal $1$\,$\sigma$ limits for point sources in the $3.6$ and $4.5$\,$\mu$m bands from the SENS-PET exposure time calculator that correspond to the maximum coverage depth in the GREATS mosaics ($250$\,hr and $200$\,hr, for the $3.6$ and $4.5$\,$\mu$m band, respectively) are $9.1$\,nJy ($29.0$\,AB) and $14.7$\,nJy ($28.5$\,AB), respectively. The $150$\,hr coverage (that we adopted  in Section \ref{sect:exptime} as a typical depth of the GREATS mosaics) corresponds to $1$\,$\sigma$ limits of $11.8$\,nJy ($28.7$\,AB) and $17.0$\,nJy ($28.3$\,AB) for the $3.6$ and $4.5$\,$\mu$m bands, respectively. The available coverage in the $5.8$ and $8.0$\,$\mu$m bands is shallower.  In the GOODS-N field, the maximum coverage is  $\sim90$\,hr, corresponding to $1$\,$\sigma$ photometric limits of $141$\,nJy ($26.0$\,AB) and $172$\,nJy ($25.8$\,AB) for the $5.8$ and $8.0$\,$\mu$m bands respectively. Meanwhile, the maximum depth over the GOODS-S field is $\sim40$\,hr, corresponding to $1$\,$\sigma$ limits of $211$\,nJy ($25.6$\,AB) and $258$\,nJy ($25.4$\,AB) in the $5.8$ and $8.0$\,$\mu$m mosaics, respectively.

\begin{figure}
\hspace{-0.5cm}\includegraphics[width=9cm]{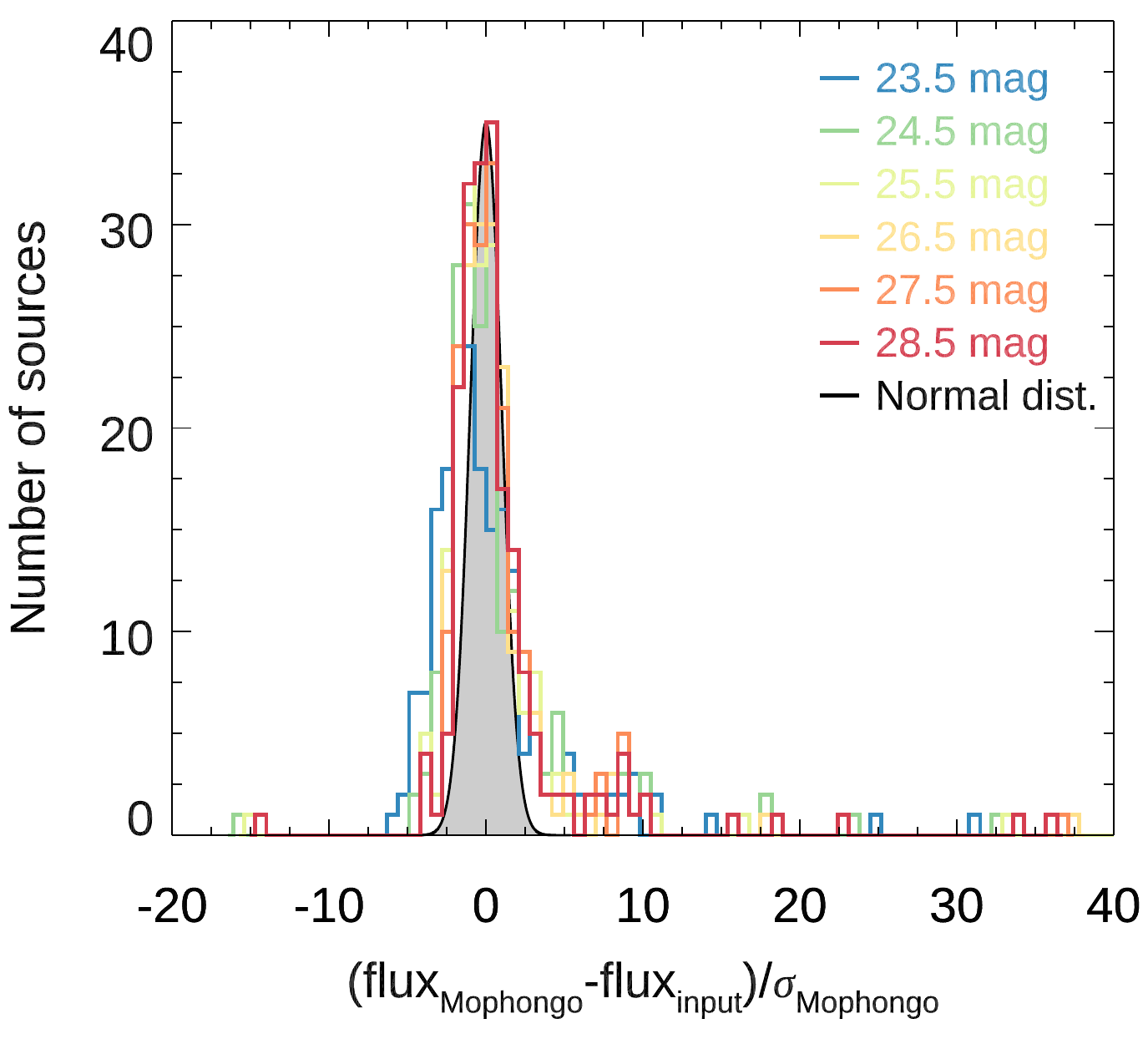}
\caption{Distribution of the differences between the flux density of synthetic sources recovered by \textsc{Mophongo} (flux$_\mathrm{\textsc{Mophongo}}$) and their nominal values (flux$_\mathrm{input}$), in units of flux density uncertainty ($\sigma_\mathrm{\textsc{Mophongo}}$). The synthetic sources were injected at random locations within one of the regions of the $3.6$\,$\mu$m GOODS-N mosaic with the deepest coverage. Each color corresponds to a specific value of the input flux density, indicated by the legend. The grey solid area with black outline corresponds to a Normal distribution with a peak value of $35$, approximately corresponding to the peak value of the histograms. Most of the measurements follow a Normal distribution with only a small fraction of sources deviating appreciably from it ($\lesssim10\%$ of the sources have differences $>5$\,$\sigma$). These results, (see also an earlier demonstration by \citealt{labbe2015}), indicate that the source confusion effects in deep IRAC imaging can be reliably mitigated even in the faintest regimes.}
 \label{fig:mophongo_test}
\end{figure}

However, the actual detection significance will depend on the impact of source confusion on the photometry for specific sources. For deep imaging with low resolution, as it is the case for the GREATS $3.6$ and $4.5$\,$\mu$m mosaics, confusion from source blending may decrease the actual sensitivity. For an instrument with a resolution similar to IRAC, \citet{franceschini1991} estimated the confusion limit to be $\sim0.6-2$\,$\mu$Jy (depending on whether the source is point-like or extended). According to the SENS-PET calculator, this limit is already exceeded in the shallowest regions of the GREATS mosaic.  However, \citet{franceschini1991} advanced the hypothesis, quantified by \citet{labbe2015}, that high-resolution imaging of the same patch of the sky can be used to effectively deblend the sources in the low-resolution image, enabling measurements down to fainter limits. Below we briefly summarize the results of \citet{labbe2015}.

\begin{deluxetable*}{cccccccc}
\tablecaption{GREATS photometry for the LBGs of \citet{bouwens2015} \label{tab:phot}}
\tablehead{\colhead{ID} & \colhead{R.A.\tablenotemark{a}} & \colhead{Dec.\tablenotemark{b}} & \colhead{$z$ bin\tablenotemark{c}} & \colhead{$3.6$\,$\mu$m\tablenotemark{d}} & \colhead{$4.5$\,$\mu$m\tablenotemark{e}} & \colhead{$5.8$\,$\mu$m\tablenotemark{f}} & \colhead{$8.0$\,$\mu$m\tablenotemark{g}}\\
 & \colhead{[hh:mm:ss.sss]}  & \colhead{[dd:pp:ss.ss]} & & \colhead{[nJy]} & \colhead{[nJy]} & \colhead{[nJy]} & \colhead{[nJy]}
}
\startdata
     GNWB-7485214158&    12:37:48.529 &     62:14:15.83 & $  4$ & $ 112.7\pm  25.8$ & $  60.3\pm  30.8$ & $-193.9\pm 252.9$ & $ 330.1\pm 347.2$  \\
     GNWB-7502914551&    12:37:50.292 &     62:14:55.12 & $  4$ & $ 158.0\pm  31.5$ & $ 125.2\pm  28.3$ & $ 189.5\pm 265.1$ & $-204.9\pm 330.7$  \\
     GNWB-7546714582&    12:37:54.674 &     62:14:58.20 & $  4$ & $ 215.6\pm  56.7$ & $ 159.1\pm  55.8$ & $ 291.7\pm 260.7$ & $-254.5\pm 460.0$  \\
     GNWB-7485914550&    12:37:48.598 &     62:14:55.07 & $  4$ & $ 132.9\pm  26.9$ & $ 102.4\pm  22.9$ & $-312.4\pm 247.2$ & $-107.2\pm 329.8$  \\
     GNWB-7575615181&    12:37:57.561 &     62:15:18.16 & $  4$ & $ 141.7\pm 102.4$ & $  98.7\pm  66.2$ & $ -59.1\pm 241.3$ & $ 857.1\pm 357.5$  \\
\enddata
\tablecomments{This Table only presents data for the first five sources. It is available in its entirety in machine-readable form. }
\tablenotetext{a}{Right Ascension of the source.}
\tablenotetext{b}{Declination of the source.}
\tablenotetext{c}{Redshift bin originally assigned to the source by the analysis of \citet{bouwens2015}.}
\tablenotetext{d}{Flux density and $1 \sigma$ uncertainty in the $3.6$\,$\mu$m band.}
\tablenotetext{e}{Flux density and $1 \sigma$ uncertainty in the $4.5$\,$\mu$m band.}
\tablenotetext{f}{Flux density and $1 \sigma$ uncertainty in the $5.8$\,$\mu$m band.}
\tablenotetext{d}{Flux density and $1 \sigma$ uncertainty in the $8.0$\,$\mu$m band.}

\end{deluxetable*}

\begin{figure*}
\begin{tabular}{c}
\hspace{-0.7cm}\includegraphics[width=18cm]{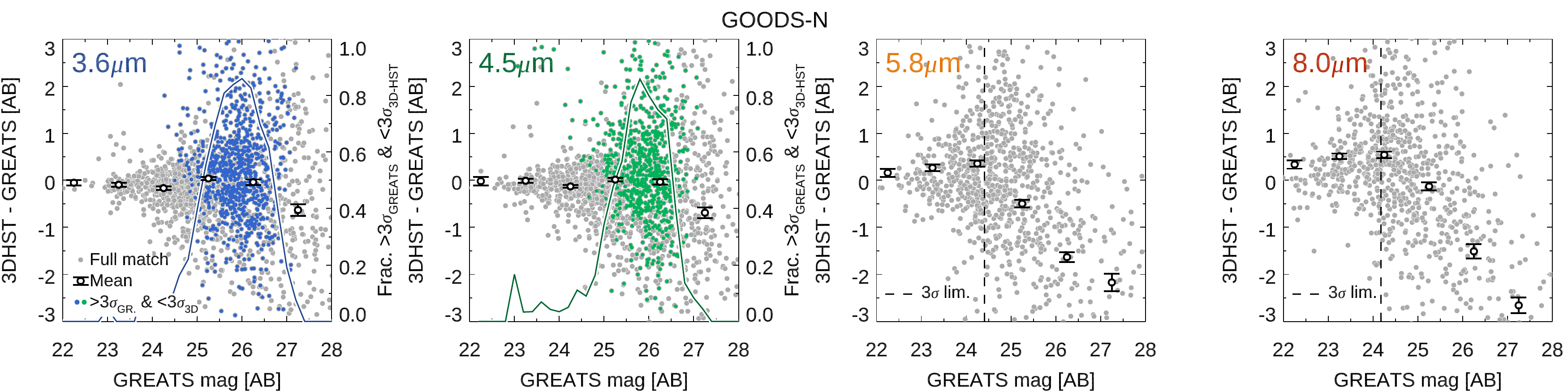}\\
\hspace{-0.7cm}\includegraphics[width=18cm]{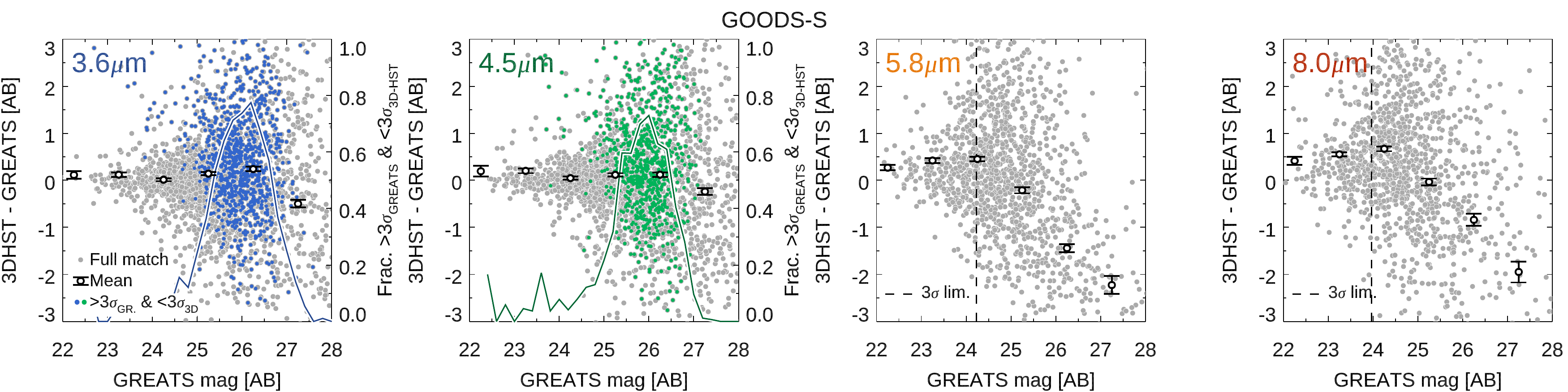}
\end{tabular}
\caption{Comparison between the IRAC photometry extracted from the GREATS mosaics and that from the 3D-HST program (\citealt{skelton2014}), for the matching LBG sample of \citet{bouwens2015}. The top row refers to sources in the GOODS-N field, while the bottom row to sources in GOODS-S. In each row, each panel corresponds to individual IRAC bands, as labelled in the top-left corners. In each panel, the grey filled circles mark the difference in magnitude of the matching sources between the two catalogs, while the black open circles with error bars the mean and the error on the mean. In the panels corresponding to the $3.6$ and $4.5$\,$\mu$m measurements, the blue and green points mark those sources which have $\ge3$\,$\sigma$ significance in the GREATS data but $<3$\,$\sigma$ in the 3D-HST catalogs, providing a first assessment of the gain in sensitivity enabled by GREATS. The solid curve corresponds to the fraction of such sources compared to the underlying population of LBGs, and it is referred to the vertical axis on the right. The dashed vertical line in the $5.8$ and $8.0$\,$\mu$m panels marks an approximate $3$\,$\sigma$ limit. No significant offsets exist with previous photometry. It is evident how GREATS mostly benefits sources around $26$\,mag, corresponding to $\sim L^*$ at $z\sim8$ (e.g., \citealt{stefanon2019}). \label{fig:phot}}
\end{figure*}

Using prior-based photometry of synthetic zero-flux sources in the GOODS-S field,  \citet{labbe2015} studied the relation between the point-source sensitivity and integration time.  The considered depths ranged from  $\sim30$\,hr to $200$\,hr, very similar to the range available for the GREATS mosaics. They found that the $1$\,$\sigma$ limit of flux density decreases approximately as the nominal $1/\sqrt{t_\mathrm{exp}}$ expected for Poisson statistics (see their Equations 3 and 4 and their Figure 11 -- the measured exponents are $-0.453$ and $-0.456$ for the $3.6$ and $4.5$\,$\mu$m band, respectively, consistent with the Poissonian $-0.5$ exponent at a $2.5$\,$\sigma$), and found no indication for any confusion limits or noise floor down to the deepest regions (\citealt{labbe2015}). However, they found that limits in flux density were consistently $10\%-30\%$ shallower than the nominal values expected from SENS-PET for the same integration times (see Figure 11 of \citealt{labbe2015}). These discrepancies could be the result of residual confusion from either sources below the detection limit in the high-resolution image or from extended  wings of the surface brightness profile, that could still could contribute noise. 

To further ascertain the impact of source confusion in the extraction of flux densities from deep IRAC images, we performed a Monte Carlo simulation, consisting of injecting synthetic point sources at random positions across a rectangular area of $\sim4\farcm1 \times 2\farcm5$ centered at [R.A., Dec.]=[$12$:$36$:$47$, $+62$:$13$:$08$] in the $3.6$\,$\mu$m GOODS-N mosaic from GREATS. This region is characterized by $\sim250$\,hr-deep coverage, one of the deepest existing for IRAC. The choice of a point-like morphology is likely not significantly impacting our results due to the small sizes of faint high-redshift sources (e.g., \citealt{shibuya2015,bouwens2021}) and the broad IRAC PSFs.

Synthetic sources with the same intrinsic brightness were added at $20$ random positions. Their flux densities were successively extracted with \textsc{Mophongo} (\citealt{labbe2006, labbe2010b, labbe2010a, labbe2013, labbe2015}).  This tool leverages the brightness profile of each source from a high-resolution image to remove all neighbouring objects within a radius of $9\farcs0$, before performing aperture photometry. Photometry was extracted adopting $1\farcs8$-diameter apertures, and corrected to total using the brightness profile of each source on the low-resolution image and the PSF reconstructed at the specific locations of each source. These steps were executed for different values of intrinsic brightness (from $23.5$\,mag to $28.5$\,mag at constant intervals of $1$\,mag). The full procedure was repeated $10$ times to improve the statistical significance. The results of this simulation are summarized in Figure \ref{fig:mophongo_test}, and show that the systematic effects from source confusion in the deep GREATS mosaics can be robustly accounted for.

In the present work, to facilitate the comparison with other surveys, all quoted flux density limits refer to the nominal values provided by the SENS-PET calculator. However, to provide a sense of the effect of source blending on the actual limits, we also present the $1$\,$\sigma$ depths estimated with the relations of \citet{labbe2015}. We find that the deepest $250$\,hr coverage in the $3.6$\,$\mu$m band corresponds to $14.1$\,nJy ($28.5$\,AB), while the $200$\,hr in the $4.5$\,$\mu$m band corresponds to $17.6$\,nJy ($28.3$\,AB). Similarly, the $150$\,hr coverage corresponds to $17.8$\,nJy ($28.3$\,AB) and $20.0$\,nJy ($28.1$\,AB) for the $3.6$ and $4.5$\,$\mu$m bands, respectively.

\section{IRAC Photometric catalog for $\sim10000$ Lyman-Break galaxies in the GOODS fields}

\label{sect:phot}

We also publicly release photometric information extracted from the GREATS mosaics for the Lyman-Break galaxy samples identified by \citet{bouwens2015} in the GOODS-N and GOODS-S fields (including ERS, XDF, HUDF091 and HUDF092). This sample is composed by a total of $9192$ sources ($3914$ sources in GOODS-N, $5278$ in GOODS-S) initially selected to have redshifts $z\sim3.5-10$ (see \citealt{bouwens2015} for details). This catalog was already adopted by the studies of \citet{debarros2019} and \citet{stefanon2021b, stefanon2021a}. We remark here that the new, deeper IRAC data provided by the GREATS mosaics could lead to  $z\lesssim3-4$ solutions to be more likely for some of the sources in the catalog. We therefore advise the interested reader to perform a full SED analysis including both \textit{HST} and IRAC photometry, if a more robust sample of LBGs is needed. Joint NIR-selected multi-wavelength catalogs incorporating the GREATS data will be released in a forthcoming paper.

Flux densities and uncertainties were extracted with \textsc{Mophongo} adopting a combination of WFC3 F125W-, F140W- and F160W-band mosaics as a high-resolution prior. Photometry was extracted adopting $1\farcs8$-diameter apertures, and corrected to total using the brightness profile and the PSF curve of growth of each source.

Table \ref{tab:phot} presents the first five entries of the photometric catalog, for reference. The full catalog is available online. In Figure \ref{fig:phot}, instead, we compare our new GREATS photometry to that of \citet{skelton2014}, who adopted the SEDS v1.2 data release (\citealt{ashby2013}) for the $3.6$ and $4.5$\,$\mu$m bands, and the GOODS \textit{Spitzer} release for the imaging at $5.8$ and $8.0$\,$\mu$m. No significant offset is evident between the two sets of photometry, in particular in the $3.6$ and $4.5$\,$\mu$m bands, where the new observations from the GREATS program were added.  It is also remarkable the gain in S/N for $\sim70-90\%$ of the sources around $\sim26$\,AB in these two bands. The marginal differences in the $5.8$ and $8.0$\,$\mu$m band photometry likely result from the combined impact of an improved background subtraction, brightness profile modelling and PSF reconstruction implemented in the version of \textsc{Mophongo} adopted to extract the photometry from the GREATS data.

\section{Key Science Enabled by GREATS}
\label{sect:science}

There has been a significant amount of progress in quantifying the properties of galaxies in the first few billion years of the universe (see e.g. \citealt{stark2016}).   Nevertheless, a large number of open questions remain which require large area, deep IR datasets. There are important, contemporary science issues that can only be well-addressed by IR observations of greater sensitivity and over greater area.  While \textit{JWST} will ultimately provide such datasets, the GREATS mosaics provide opportunities to work with such IR data \textit{now}. The purpose of this section is to present a few of the main science cases that will set the stage for advancing the field and providing focus for future \textit{JWST} observations.

\subsection{Stellar mass estimates at $z>7$}

Characterizing the mass buildup of the first galaxies provides important constraints to galaxy formation models (e.g. \citealt{behroozi2013, behroozi2019, tacchella2018}). A number of works have estimated the assembly of  stellar mass ($M_\star$) with redshift down to the first $\sim1$\,Gyr of cosmic history ($z\gtrsim6$ - e.g., \citealt{gonzalez2014, duncan2014, grazian2015, song2016, stefanon2017b, davidzon2017,  bhatawdekar2019, kikuchihara2020}). However, the area and sensitivity of the pre-GREATS IRAC coverage only allows individual detections of a handful of galaxies at $z\sim8$, with similar numbers from stacking analysis. To robustly measure the stellar mass assembly at $z\sim8-10$, significant samples of $z>8$ LBGs individually detected in IRAC bands are required. These large rest-frame optical samples can significantly reduce the impact of uncertainties in SFHs or dust content of galaxies relative to studies executed with only rest-frame UV data. The GREATS mosaics provide nominal $\ge 4$\,$\sigma $ individual detections at $3.6$\,$\mu$m for $\sim50\%$ of the sources at $z\sim8$ with $L\ge0.4L^*$  over CANDELS GOODS-S and GOODS-N. As GREATS coverage only exists over $\sim50\%$ of the CANDELS GOODS footprint, this is effectively equivalent to our detecting $100\%$ of the galaxies lying in areas with ultra-deep coverage from IRAC. This addresses the need for large samples, but the impact of dense fields on the sensitivity estimates (see the discussion in Sect.~\ref{sect:depth}) needs to be kept in mind.

\subsection{Nebular emission at $z>6$}
There is increasing amount of evidence indicating that galaxies at $z\gtrsim 4$ are characterized by strong nebular emission at rest-frame optical, with equivalent width EW([\ion{O}{3}]+$H\beta$) in excess of 800\,\AA\   (\citealt{fumagalli2012, stark2013, smit2014, smit2015, faisst2016, rasappu2015,  khostovan2016, stefanon2017c, stefanon2019, debarros2019, lam2019, bowler2020}). These fascinating results suggest that high-redshift galaxies likely have high specific SFR (sSFR$\gtrsim 10 $\,Gyr$^{-1}$) and young ages (few tens $\times$\,Myr to few hundreds $\times$\,Myr). Such impressive results have been obtained by characterizing the strong variations of the  $[3.6]-[4.5]$ color with redshift. The flux changes seen reflect the contribution of the main emission lines  (e.g., [\ion{O}{3}], $H\beta$, $H\alpha$, [\ion{N}{2}]) entering and leaving the $3.6$ and $4.5$\,$\mu$m bands depending on the redshift (see e.g., \citealt{labbe2013, smit2014, faisst2016}). The depth of IRAC observations prior to GREATS however limited the exploration for such lines in sources beyond $z\sim6$.  GREATS enables the investigations of the occurrence of such lines to finally be extended to $z>6$. Furthermore, recent studies have shown that the existence of strong rest-frame optical lines can substantially, and incorrectly, boost the estimates of stellar mass (e.g., \citealt{stark2013}) if not accounted for. Therefore robust estimates of the emission line strength will also improve stellar mass estimates. GREATS has already been used to demonstrate the gains that will be made with this datatset. This can be seen in the recent paper by  \citet{debarros2019} where new [\ion{O}{3}]+H$\beta$ measurements at $z\sim8$ are made based on the GREATS IRAC $3.6$ and $4.5$\,$\mu$m mosaics.

\subsection{The rise of passive galaxies}
A major uncertainty in current models of galaxy evolution is when and how passive galaxies start to appear in the Universe. There have already been some exciting selections of evolved, red galaxies  identified at $z\sim 4-6$ (e.g., \citealt{guo2012, caputi2012, straatman2014, stefanon2015}), with a record-holder at $z\sim9$ (\citealt{hashimoto2018}). Only a handful of such passive galaxies are spectroscopically confirmed (\citealt{marsan2017, glazebrook2017, schreiber2018, hashimoto2018, forrest2020}). Due to their red colors, such galaxies can remain completely undetected in current \textit{HST} datasets, making IRAC the only instrument currently able to detect the most extreme sources. Due to the limited depth and area of current IRAC imaging, current searches for passive systems at $z > 4$ have so far exclusively focused on the high-mass end ($M_\star\sim10^{10.5-11}$\,$M_\odot$). Ultradeep IRAC observations like those from GREATS will enable us to access lower stellar masses $M_\star\lesssim 10^{9.5-10}$\,$M_\odot$ where the likelihood of finding such sources may increase.

\subsection{SFR-$M_\star$ at $z>7$}
A growing number of observations have identified a tight relation ($\sigma\lesssim0.3$\,dex - (\citealt{daddi2007, whitaker2012, schreiber2015,kurczynski2016}) between the SFR and $M_\star$ (the so-called \textit{main sequence} of star-forming galaxies - \citealt{noeske2007}) up to  $z\sim6-7$ (e.g., \citealt{labbe2010b,bouwens2012, stark2013,  gonzalez2014, duncan2014,  steinhardt2014, salmon2015,  song2016, jiang2016, santini2017, kartheik2018, pearson2018, khusanova2020, faisst2019}). Indeed, a correlation between the SFR and $M_\star$ is predicted by models (e.g., \citealt{finlator2007,finlator2011, finlator2018, dekel2009,dave2011, lu2014, mancuso2016b, wilkins2017, ma2018, ceverino2018, rosdahl2018, tacchella2018}). The scatter, slope and normalization constrain the SFH, feedback mechanisms and cold gas accretion. However, limitations in the estimates of stellar mass from current rest-frame UV-selected LBGs have only allowed a tentative determination of this relation at high redshifts ($z\sim7$ - \citealt{labbe2010b}). The samples have been further limited to the brightest sources. The lower stellar mass limits enabled by GREATS ($M_\star\sim 10^{9.5}$\,$M_\odot$ for $z\lesssim8-9$) will allow us to include $\sim 5\times$ more sources at $z>4$, and most importantly, with more robust stellar mass estimates.

\section{Public Data Release}
\label{sect:dr}

The data release accompanying this paper consists of the reduced images of all ultradeep IRAC observations over the GOODS-North and GOODS-South fields. We enhance this data release with flux density estimates in the four IRAC bands for all the Lyman-Break galaxies in the GOODS fields identified by \citet{bouwens2015} at $z\sim3.5-10$.

Specifically, this data release includes the following:

\begin{enumerate}
\item Science images and coverage maps in the $3.6$, $4.5$, $5.8$ and $8.0$\,$\mu$m bands. Our reduction uses the same tangent point as CANDELS on pixel scales of $0\farcs3$\,pixel$^{-1}$, so the IRAC maps can be easily rebinned and registered to \textit{HST}/WFC3 data.
\item Reduced images of all individual 845 AORs, drizzled onto the same grid, which may be useful to study the reliability or variability of sources.
\item Template PSFs and spatial maps of the weights and position angles of each AOR, allowing the reconstruction of the PSF at arbitrary locations. A python code to reconstruct the PSF at the desired location within the mosaic is also provided.
\item A catalog containing the position, flux density and uncertainty estimates in the four IRAC bands for the $9192$ sources identified by \citet{bouwens2015} in the CANDELS/GOODS fields as candidate Lyman-Break galaxies consistent with redshifts in the range $z\sim3.5-10$ (\citealt{bouwens2015}).
\end{enumerate}

 These data products are publicly available through the IRSA website. The units of the science images are MJy/sr. Equivalently, flux densities can be obtained by multiplying the image pixel values by $2.1154\mu$\,Jy\,DN$^{-1}$, corresponding to an image AB zeropoint of 23.0865\,mag. Coverage maps are in units of seconds based on the warm mission observations; the higher sensitivity of the cryogenic observations is dealt with by scaling the cryogenic exposure times by a factor $1.7$.

\section{Summary and Conclusions}
\label{sect:conclusions}

Our mosaics include $\sim610$\,hr of new observations per band from the GOODS Re-ionization Era wide-Area Treasury from Spitzer (GREATS, PI: I. Labb\'e) program. Remarkably, these new GREATS observations enable a significant transformation of all the available data on GOODS-S and GOODS-N into a uniquely deep and wide imaging resource. The GREATS Mosaics are the result of combining $2132$\,hr of observation in the $3.6$ and $4.5$\,$\mu$m IRAC band, and  $488$\,hr in the $5.8$ and $8.0$\,$\mu$m bands. GREATS data extend the ultradeep coverage in the $3.6$ and $4.5$\,$\mu$m bands with $>150$\,hr of deep data (corresponding to a $1$\,$\sigma$ sensitivity for point sources of $\sim12$\,nJy in the $3.6$\,$\mu$m band - SENS-PET) across $\sim150$\,arcmin$^2$ ($\sim1/2$ the total area of the GOODS fields). This area is $5\times$ the area currently covered by the similarly-deep previous dataset IUDF (\citealt{labbe2015}). The GREATS mosaics reach an impressive $250$\,hr coverage in a small $\sim5-10$\,arcmin$^2$ region in each field in the $3.6$ and $4.5$\,$\mu$m bands. Through accurate planning of the new GREATS observations, there is a good match between the depth in the $4.5$\,$\mu$m coverage and the $3.6$\,$\mu$m coverage over the two GOODS fields. Specifically, $\sim95\%$ of the area in the GOODS fields  with $>75$\,hr depth in the $4.5$\,$\mu$m band has $>90\%$ of that integration time in the $3.6$\,$\mu$m band (best seen in Figure \ref{fig:exptime_comparison}). The nominal sensitivity of the IRAC coverage is a close match to that available with WFC3/IR over the CANDELS Deep regions, enabling  $\sim4$\,$\sigma$ detections for $\sim$half of the sample of $L>0.4L^*$ galaxies at $z\sim8$.  This allows for the characterization of optical emission lines and the much more reliable estimate of stellar masses for a substantial samples of galaxies at $z\sim7-10$. Added knowledge of these aspects are important to fully prepare for science with \textit{JWST}.

Our public release includes for each field and band the science mosaics and corresponding coverage maps. In addition, the release also includes PSF maps which take into account the complex spatial variation of the survey geometry. These PSF maps are an essential resource for optimally mitigating the impact of source blending. The science frames are calibrated to an AB zeropoint of $23.0865$mag, while the coverage maps have units of seconds and include a $1.7\times$ upscaling to account for the higher sensitivity achieved during the cryogenic mission. The mosaics are characterized by a quite uniform PSF FWHM$\sim1\farcs49-1\farcs52$  at $3.6$ and $4.5$\,$\mu$m ($\sim 1\farcs7-2\farcs0$ in the $5.8$ and $8.0$\,$\mu$m bands, respectively) across the fields. The uniformity of the PSF FWHM provides  consistent image quality. The mosaics have the tangent point of the CANDELS/GOODS fields with a pixel scale of $0\farcs3$\,pixel$^{-1}$. Finally, we also release the photometry in the four IRAC bands, obtained after removing the contamination from neighbours, for $9192$ candidate Lyman-Break galaxies at $z\sim3.5-10$ identified by \citet{bouwens2015} over the GOODS-N and GOODS-S fields.

The science investigations enabled by the GREATS Mosaic will both enhance the field before \textit{JWST} and also provide a framework for optimizing the early observations made with \textit{JWST}.  This will be the deepest and largest-area IR dataset at $3-5$\,$\mu$m until \textit{JWST} begins its observations.

\acknowledgements
The authors would like to thank the two referees for their careful reading of the manuscript and for providing constructive comments that helped improving the quality of the paper. MS and RJB acknowledge support from TOP grant TOP1.16.057. PAO acknowledges support from the Swiss National Science Foundation through the SNSF Professorship grant 190079 `Galaxy Build-up at Cosmic Dawn'. The Cosmic Dawn Center (DAWN) is funded by the Danish National Research Foundation under grant No.\ 140.  We also acknowledge the support of NASA grants HSTAR-13252, HST-GO-13872, HST-GO-13792, and NWO grant 600.065.140.11N211 (vrij competitie). GDI acknowledges support for GREATS under RSA No. 1525754. This paper utilizes observations obtained with the NASA/ESA \textit{Hubble Space Telescope}, retrieved from the Mikulski Archive for Space Telescopes (MAST) at the Space Telescope Science Institute (STScI). STScI is operated by the Association of Universities for Research in Astronomy, Inc. under NASA contract NAS 5-26555. This work is based [in part] on observations made with the \textit{Spitzer Space Telescope}, which was operated by the Jet Propulsion Laboratory, California Institute of Technology under a contract with NASA. Support for this work was provideed by NASA through an award issued by JPL/Caltech. 

\bibliographystyle{apj}

\bibliography{mybib}

\end{document}